\documentclass[useAMS,usenatbib]{mn2e}
\usepackage{newtxtext,newtxmath}
\usepackage{multirow}
\usepackage{times}
\usepackage{graphics,epsfig,txfonts}
\usepackage{textcomp}
\usepackage{subfigure}

\usepackage{enumerate}
\usepackage[T1]{fontenc}
\usepackage{ae,aecompl}
\usepackage[usenames, dvipsnames]{color}
\usepackage[flushleft]{threeparttable}

\usepackage{natbib}

\usepackage{graphicx}	
\usepackage{mathtools}
\usepackage{amsmath}	
\usepackage{amssymb}	
\usepackage{pdflscape}

\title[SMC Radio pulsar population]{The radio pulsar population of the Small Magellanic Cloud}

\author[N. Titus et al.]{
N. Titus$^{1,2}$\thanks{E-mail: naomi@saao.ac.za},
S. Toonen$^{3}$,
V. A. McBride$^{1,2,4}$,
B. W. Stappers$^{5}$,
D. A. H.Buckley$^{2}$,
\newauthor
L. Levin$^{5}$
\\
$^{1}$Department of Astronomy, University of Cape Town, Private Bag X3, Rondebosch, 7701, South Africa\\
$^{2}$South African Astronomical Observatory, PO Box 9, Observatory, 7935, South Africa\\
$^{3}$Institute of Gravitational Wave Astronomy and School of Physics and Astronomy, University of Birmingham, Birmingham, B15 2TT, United Kingdom \\
$^{4}$IAU Office of Astronomy for Development, Cape Town, Observatory, 7935, South Africa\\
$^{5}$Jodrell Bank Centre for Astrophysics, School of Physics and Astronomy, University of Manchester, Manchester M13 9PL, UK\\
}

\date{Accepted 2020 March 01. Received 2020 February 24; in original form 2020 January 09}

\pubyear{2017}

\begin{document}
\label{firstpage}
\pagerange{\pageref{firstpage}--\pageref{lastpage}}
\maketitle

\begin{abstract}
We model the present day, observable, normal radio pulsar population of the Small Magellanic Cloud (SMC). The pulsars are generated with \texttt{SeBa}, a binary population synthesis code that evolves binaries and the constituent stellar objects up to remnant formation and beyond. We define radio pulsars by selecting neutron stars that satisfy a selection of criteria defined by Galactic pulsars, and apply the detection thresholds of previous and future SMC pulsar surveys.  The number of synthesised and recovered pulsars are exceptionally sensitive to the assumed star formation history and applied radio luminosity model, but is not affected extensively by the assumed common envelope model, metallicity, and neutron star kick velocity distribution.   We estimate that the SMC formed (1.6\,$\pm$\,0.3)\,$\times$10\textsuperscript{4} normal pulsars during the last 100\,Myrs.  We study which pulsars could have been observed by the Parkes multibeam survey of the SMC, by applying the survey's specific selection effects, and recover  4.0\,$\pm$\,0.8 synthetic pulsars. This is in agreement with their five observed pulsars.   We also apply a proposed MeerKAT configuration for the upcoming SMC survey, and predict that the MeerKAT survey will detect 17.2\,$\pm$\,2.5 pulsars.    
\end{abstract}




\section{Introduction}

The Small Magellanic Cloud (SMC) is an excellent candidate to study a variety of stellar populations, since it is located out of the Galactic plane, providing minimal extinction, yet being close enough to resolve stars across the whole galaxy.  Another advantage is the known distance to the SMC, necessary for the calculation of luminosities that lead to accurate classifications of stellar populations.  The SMC is on average at a distance of 60\,kpc \citep{graczyk2014}, however it is believed to have a line of sight depth of $\sim$10\,kpc \citep{crowl2001}.  This uncertainty introduces a minimal uncertainty of a few percent for the calculated luminosities.  For reasons such as these the SMC has been surveyed extensively at all wavelengths, producing representative samples of the various underlying stellar populations, within the constraints of the surveys.  

The SMC contains a variety of neutron star (NS) populations that highlight various evolutionary stages in the lives of massive stars and massive binaries.  The radio pulsar population of the SMC is of particular interest, since only seven pulsars have been discovered by previous surveys \citep{mcconnell1991,crawford2001,manchester2006,titus2019}.  Although these surveys optimised survey sensitivity and coverage, they were only sensitive enough to detect the brightest pulsars due to the large distance to the SMC.  In contrast to the small radio pulsar population, the SMC contains 120 high mass X-ray binaries (HMXBs) of which $\sim$60 are confirmed NS X-ray pulsars \citep{haberl2016}.  The previous X-ray surveys of the SMC (e.g.
\citealt{yokogawa2000}; \citealt{haberl2004}; \citealt{shtykovskiy2005}; \citealt{sturm2013}) were sensitive enough to probe accreting NSs in quiescence and outburst, unveiling a representative sample of the underlying X-ray binary population.  Population synthesis studies can establish if the relatively small numbers of radio pulsars are an underlying property of the NS population of the SMC, or to what extent observational biases and constraints are affecting the observed radio pulsar population.  The population synthesis results can be compared to previous radio pulsar surveys and population studies of the SMC \citep{ridley2010}, as well as predict the number of pulsars that future surveys should detect, like MeerKAT or SKA. 

The star formation history of the SMC plays a critical role when simulating the observable normal radio pulsar\footnote{Normal pulsars refer to non-accreting, non-millisecond pulsars.} population.  However, it is only the most recent star formation episodes (SFEs) that will contribute to the observable pulsar population, since the average lifetime of an isolated radio pulsar is $\sim$100\,Myr \citep{faucher2006,ridley2010}.  During every SFE a population of massive stars will be born. Multiplicity studies of massive stars indicate that most massive stars are born in binaries \citep{kobulnicky2007,kouwenhoven2007,sana2012,sana2014,moe2017}. Binarity will impact the evolution of the constituent stars if they exchange material.  An isolated massive star will evolve to the giant phase, and produce a NS during a core collapse supernova (Type Ib/c; \citealt{heger2003}).  However massive stars in close binaries exchange material, forming at least one helium star, leading to a core collapse supernova and likely the formation of a NS \citep{yoon2010}.  NSs form via different evolutionary paths, and are born with a range of birth spin periods and magnetic fields. Moreover, if the NS is not accreting matter from a companion, it can enter a radio pulsar phase.  The radio pulsar mechanism can switch on if the NS has a fast enough spin period, and a sufficiently strong surface magnetic field, resulting in highly beamed, coherent radio emission from the magnetic poles.

Given the prevalence of binarity in massive star populations, we conducted binary population synthesis studies to best simulate the events leading up to the formation of radio pulsars as well as the evolution of the pulsars.  In this paper we present our population synthesis results of the SMC radio pulsar population and predict the number of pulsars MeerKAT will detect.  In Section 2, we describe the simulation parameters and define a radio pulsar within the simulations.  In Section 3 we present the results, and discuss their implications in Section 4.  In Section 5 we draw our conclusions.

\section{Method}

We utilised the binary population synthesis code \texttt{SeBa} \citep{portegies1996,toonen2012,toonen2013} to model the evolution of massive binaries and the formation of isolated and radio pulsars in binaries.  \texttt{SeBa} evolves the binaries from the zero age main sequence to remnant formation.  Beyond remnant formation \texttt{SeBa} continues to evolve the individual remnants, as well as the binary interaction for the remaining binaries.  The binary interaction accounts for mass loss, mass transfer, angular momentum loss, common-envelope (CE) phase, magnetic braking, gravitational radiation, and disruption, which is applied at every time step with the relevant recipes (see \citealt{portegies1996}; \citealt{portegies1998}).  For this study pertaining to radio pulsars, we limit our selection of sources to detached and disrupted binaries containing a NS.  A detached binary implies there is no mass transfer via Roche lobe overflow (RLOF), however mass transfer via wind accretion is still allowed.   A disrupted binary is defined as a system that is no longer gravitationally bound as a consequence of a supernova explosion.


\subsection{Initial Conditions of default model}\label{sec:initial_conditions}

We simulated 3 million binaries with \texttt{SeBa} to ensure a statistically significant population of radio pulsars.  Here follows a list of the initial conditions required for our default model to simulate the population of binaries.  For the CE evolution we used the $\gamma\alpha$ model (see Equation~\ref{eqn:ce}, Section~\ref{sec:ce}) with a combined CE parameter $\alpha_{CE}\lambda$\,=\,0.25 \citep{toonen2013}, where $\gamma$\,=\,1.75 \citep{portegies1998,toonen2012,vanhaaften2015}.   The initial primary masses ($M_1$\,stars with the potential of forming NSs) are drawn from a Kroupa initial mass function (IMF, \citealt{kroupa2001}) with exponent 2.3 for masses between 4.4 and 40\,M$_\odot$.  Both mass limits were chosen to include all possible NS formation scenarios.  The lower mass limit was calculated by simulating 500\,000 binaries and determining the minimum primary mass required to form a NS.  The secondary masses ($M_2$) are drawn from a uniform mass ratio distribution such that 0\,<\,$\frac{M_2}{M_1}$\,<1 \citep{raghavan2010,duchene2013}.  The semi-major axis follows a uniform distribution in $\log(a)$ \citep{abt1983} with a maximum cut off value of 1$\times$10$^6$\,R$_\odot$.  The initial eccentricity values are selected from a thermal distribution \citep{heggie1975} ranging between zero and one.  The natal NS kick velocity distribution is modelled by \citet{verbunt2017}, a combination of two Maxwellians with a mean of 75\,km\,s\textsuperscript{-1} and 316\,km\,s\textsuperscript{-1}.  For the birth NS spin period, we assume a normal distribution with a mean of 300\,ms and a standard deviation of 150\,ms \citep{faucher2006}, while the birth magnetic field in units of G is drawn from a lognormal distribution with a mean and standard deviation of 12.65 and 0.55 respectively \citep{faucher2006}.  The NS kick velocity, natal spin period and magnetic field distributions were determined from Galactic pulsars.  We set the metallicity (Z) to 0.004, and assume that all NSs are born from binary stars (i.e. a 100\% binary fraction), and can either remain in a binary following the supernova (pulsars in binaries) or alternatively be disrupted forming a population of isolated radio pulsars.

\subsection{Sensitivity to initial conditions}

To assess the systematic uncertainty of sensitivity to the initial conditions we explore the effect various parameters and models have on the number of formed and detected radio pulsars.  We vary the metallicity, CE model, combined CE parameter ($\alpha_{CE}\lambda$), and the NS kick velocity distribution.  For every combination of parameters and models we simulate 3 million binaries.  A summary of all the simulations can be found in Table~\ref{tab:parameters}.   

\begin{table}
\centering
\caption{Summary of all the independent \texttt{SeBa} simulations with their corresponding parameters and models.  The default model is printed in bold.}
\label{tab:parameters}
\begin{tabular}{lllll}
\hline
Model & Z & CE model & $\alpha_{CE}\lambda$ & NS kick velocity \\
\hline
\hline
V4CE2 &  0.004 & $\gamma\alpha$ or $\alpha\alpha$  & 2.00 & Verbunt et al. (2017)$^a$ \\
H4CE2 & 0.004 & $\gamma\alpha$ or $\alpha\alpha$ & 2.00 & Hobbs et al. (2005)$^b$ \\
\textbf{V4CE25} & \textbf{0.004} & {$\pmb{\gamma\alpha}$} or \textbf{$\alpha\alpha$} & \textbf{0.25} & \textbf{Verbunt et al. (2017)} \\
H4CE25 & 0.004 & $\gamma\alpha$ or $\alpha\alpha$ & 0.25 & Hobbs et al. (2005) \\
0.1V4CE2 &	 0.004 & $\gamma\alpha$ or $\alpha\alpha$ & 2.00 & 0.1\,$\times$\, Verbunt et al. (2017)$^c$ \\
0.1V4CE25 & 0.004 & $\gamma\alpha$ or $\alpha\alpha$ & 0.25 & 0.1\,$\times$\, Verbunt et al. (2017) \\
\hline
V8CE2 & 0.008 & $\gamma\alpha$ or $\alpha\alpha$ & 2.00 & Verbunt et al. (2017) \\
H8CE2 & 0.008 & $\gamma\alpha$ or $\alpha\alpha$ & 2.00 & Hobbs et al. (2005) \\
V8CE25 & 0.008 & $\gamma\alpha$ or $\alpha\alpha$ & 0.25 & Verbunt et al. (2017) \\
H8CE25 & 0.008 & $\gamma\alpha$ or $\alpha\alpha$ & 0.25 & Hobbs et al. (2005) \\
0.1V8CE2 & 0.008 & $\gamma\alpha$ or $\alpha\alpha$ & 2.00 & 0.1\,$\times$\, Verbunt et al. (2017) \\
0.1V8CE25 & 0.008 & $\gamma\alpha$ or $\alpha\alpha$ & 0.25 & 0.1\,$\times$\, Verbunt et al. (2017) \\
\hline
\end{tabular}
\begin{tablenotes}
\item$^a$\citet{verbunt2017}: a combination of two Maxwellians with a mean of 75\,km\,s\textsuperscript{-1} and 316\,km\,s\textsuperscript{-1}.
\item$^a$\citet{hobbs2005}: a single Maxwellian with mean of 265\,kms\textsuperscript{-1}.
\item$^c$0.1\,$\times$\, \citet{verbunt2017}: reducing the NS kick velocity drawn from \citet{verbunt2017} by a factor of 10.
\end{tablenotes}
\end{table}

\subsubsection{Common-envelope model}\label{sec:ce}

The CE phase is a short-lived phase in the life of a binary that occurs when an evolved donor extends beyond its Roche lobe, enabling unstable mass transfer leading to the formation of a CE (see \citealt{ivanova2013}, and references therein).  During the CE phase the binary orbit shrinks as angular momentum is transferred to the envelope, expelling material from the binary.  Although the CE phase is critical for the formation of compact binaries, the processes governing the CE phase are poorly understood.\\
\\
For population studies the CE evolution is typically modelled by a parametric model such as the $\alpha$\,\textendash\,formalism that is based on the binary's energy budget \citep{paczynski1976,webbink1984,dekool1987,livio1988,dekool1990}:

\begin{equation}\label{eqn:ce}
    E_{bin} \text{ = } \frac{GM_dM_c}{\lambda R} \text{ = } \alpha_{CE}E_{orbit},
\end{equation}

\noindent where $G$ is Newton's gravitational constant, $M_d$ the mass of the donor, $M_c$ the mass of its core, $R$ the donor's radius, while $\lambda$ and $E_{bin}$ refers to the structure parameter and binding energy of the envelope respectively.  The orbital energy ($E_{orbit}$) is the source that unbinds the envelope with an efficiency of $\alpha_{CE}$.  This formalism is referred to as the $\alpha\alpha$ model in the paper.  For the $\alpha\alpha$ model we set $\alpha_{CE}\lambda$\,=\,0.25 or $\alpha_{CE}\lambda$\,=\,2.  The first was determined for systems that go through a single CE phase, forming white dwarf main sequence binaries \citep{zorotovic2010,zorotovic2014,toonen2013, camacho2014}, while the second relates to systems subjected to two CE phases leading to the formation of double white dwarf binaries (Nelemans et al. 2000, 2001b).\\
\\
Another CE formalism is the $\gamma$\,\textendash\,formalism that relates to angular momentum conservation of the binary \citep{nelemans2000}:  

\begin{equation}
    \frac{J_{init} - J_{final}}{J_{init}} =\text{ = } \gamma\frac{\Delta M_d}{M_d + M_a},
\end{equation}

\noindent where $J_{init}$ and $J_{final}$ are the angular momentum of the binary prior to and following the CE phase respectively, while $M_a$ refers to the mass of the companion.  In our models we set $\gamma$\,=\,1.75. This is applied unless the binary contains a compact object or the CE phase is triggered by a tidal instability rather than dynamically unstable Roche lobe overflow (see \citealt{toonen2013}).  When this is not the case we implement the $\alpha\alpha$ CE model where $\alpha_{CE}\lambda$\,=\,0.25 or $\alpha_{CE}\lambda$\,=\,2 \citep{nelemans2001b}.  We refer to this as the $\gamma\alpha$ model.  For our default model we assume the $\gamma\alpha$ formalism and set $\alpha_{CE}\lambda$\,=\,0.25, since we are expecting the majority of the pulsars in binaries to have main sequence companions.

\subsubsection{Neutron star kick velocities}

A high mass star ending its life with a core-collapse (CC) supernova will form a NS, and impart a kick to the newly formed remnant.  The mechanism responsible for NS kicks are not well defined, nonetheless they are thought to be related to asymmetric mass loss during SNe or, alternatively to anisotropic neutrino emission \citep{lai2001,hobbs2005,janka2012}.  NSs can also receive a "Blaauw-kick" \citep{blaauw1961} when the progenitor experiences an abrupt mass loss phase, however these kicks are limited by close binary interactions that expel the donor's envelope prior to the SN \citep{huang1963,tutukov1973,leonard1994}.  Particularly, NS kicks are implied by the high spatial velocity of radio pulsars \citep{cordes1993,hobbs2005,lyne1994,verbunt2017} when compared to their progenitors \citep{gunn1970}.\\
\\
For this paper we adopt three different NS kick velocity models.  Our default model draws arbitrary velocities from the natal NS kick velocity distribution of \citet{verbunt2017}, which was determined from radio pulsar parallaxes and proper motions.  The distribution is composed of a combination of two Maxwellians with $\sigma$\,=\,75\,km\,s\textsuperscript{-1} and $\sigma$\,=\,316\,km\,s\textsuperscript{-1}.  The bimodal distribution is linked to the reduced NS kick velocity of small iron core NSs and NSs that formed via electron capture SNe a \citep{podsiadlowski2004,heuvel2004,knigge2011,janka2012,tauris2015,bray2016}.   Studies of X-ray binaries found that these NSs can have extremely low NS kick velocities \citep{pfahl2002,heuvel2004,linden2009,antoniou2010}.  To account for such low velocities we construct a second model by reducing the velocities drawn from the \citet{verbunt2017} distribution by a factor of 10.  For the third model we consider the distribution of \citet{hobbs2005} which is a single Maxwellian with a one-dimensional (1D) root mean square (rms) of $\sigma$\,=\,265\,km\,s\textsuperscript{-1} inferred from the proper motion of radio pulsars. 

\subsection{Radio pulsar models}\label{sec:obs}

A radio pulsar is modelled as a rotating magnetic dipole in a vacuum, radiating energy through dipole spin-down emission.  The energy loss results in the generation of a magnetospheric plasma and consequently the acceleration of charged particles producing radio emission (see \citealt{gurevich2007} and references therein).  The majority of all known radio pulsars have spin periods ($P$) less than 10\,s, and period derivatives \textit{(\.P)} between 10$^{-22}$ and 10$^{-10}$\,ss\textsuperscript{-1}.  The \textit{\.P} of the NS is dependent on its surface magnetic field ($B$), moment of inertia ($I$), radius ($R$), and $P$:

\begin{equation}
    \text{\textit{\.P}} \text{ = } \frac{8\pi^2R^6}{3c^3I}\frac{B^2}{P},
\end{equation}

\noindent where $c$ is the speed of light.  As such we define a simulated radio pulsar as a NS within the $P$\,\textendash\,\textit{\.P} parameter space of the observed pulsar population.

\subsubsection{Magnetic field decay}

When studying the  $P$\,\textendash\,\textit{\.P} diagram young radio pulsars appear to have stronger magnetic fields than older pulsars.  As a result \citet{gunn1970} proposed spontaneous magnetic field decay as a possible mechanism for the magnetic field evolution of isolated radio pulsars.  It has been a controversial topic and to date no consensus has been reached.  Many studies have not found convincing evidence for magnetic field decay \citep{bhattacharya1992,lorimer1997,leeuwen2004,gonthier2004,faucher2006,johnston2017}, while other studies argue for spontaneous magnetic field decay \citep{gullon2014,gullon2015,igoshev2015,cieslar2020}.  For this work we impose Ohmic decay \citep{cumming2004,geppert2009} that relates to NS crust properties of non-accreting neutron stars, and is represented by the exponential function:

\begin{equation}\label{eqn:decay}
    B\left(t\right) \text{ = } B_0\text{\,}e^{-\frac{t}{\tau}},
\end{equation}

\noindent where $t$ is time, $B_0$ is the initial (natal) magnetic field, and $\tau$\,=\,100\,Myr is the decay time scale \citep{igoshev2015}. 



\subsubsection{Radio emission mechanism}

The acceleration of electron-positron pairs is crucial for the generation of radio pulsar emission.  The radiation terminates when the pulsar's rotating magnetosphere no longer generates the potential difference required for pair production \citep{sturrock1971,chen1993}.  When the emission terminates the radio pulsar becomes radio quiet. This corresponds to a death line on the $P$\,\textendash\,\textit{\.P} diagram beyond which the pulsar mechanism switches off, and consequently the pulsar is no longer observed.  \citet{bhattacharya1992} calculated a theoretical death line of the form

\begin{equation}
    \label{eqn:death}
    \frac{B}{P^2} \text{ = } 0.17\times 10^{12}\text{ G\,s\textsuperscript{-1}},
\end{equation}

\noindent where $B$ is the NS's surface magnetic field, and $P$ the NS's spin period.  This relation is well supported by the distribution of known pulsars on the $P$\,\textendash\,\textit{\.P} diagram \citep{faucher2006}.  For our synthetic pulsars we use Equation~\ref{eqn:death} to distinguish between active and radio quiet pulsars.

\subsubsection{Observational Effects}

The most important observational effects that can hinder the detection of a radio pulsar in the SMC are the pulsar's beaming fraction and brightness.  Previous work \citep{lyne1975,vivekanand1981,proszynski1984} found that the radio luminosity for Galactic pulsars has a power law dependence on $P$ and \textit{\.P}.  Considering a power law of the form

\begin{equation}
    \label{eqn:lum}
    \log\left(L\right) \text{ = } \log\left(L_0P^{\epsilon_P}\textit{\text{\.P}}^{\epsilon_\text{\.P}}_{15}\right) \text{ + } L_\mathrm{corr}, 
\end{equation}

\noindent where $L_0$\,=\,0.18\,mJy\,kpc\textsuperscript{2}.  The units of $P$ is seconds, while \textit{\.P} has units of $10^{-15}$\,ss\textsuperscript{-1}. The two exponents relating to $P$ and \textit{\.P} are determined empirically, with values of $\epsilon_P$\,=\,-1.5 and $\epsilon_\text{\.P}$\,=\,0.5 \citep{lorimer1993,faucher2006,bates2014}.  The radio luminosity of a pulsar may be affected by physical variations in the modelled luminosity, distance uncertainties, as well as viewing geometries.  To account for these variations the dithering term, $L_\mathrm{corr}$ is added.  $L_\mathrm{corr}$ is modelled by a normal distribution, centred at zero, with $\sigma_{L_\mathrm{corr}}$\,=\,0.8 \citep{lorimer1993} determined empirically.  Once $L$ is calculated the flux density ($S$) for each pulsar is given by dividing $L$ with $d^2$, where $d$\,=\,60\,kpc is the average distance to the SMC.  This conversion from luminosity does not account for any beaming or geometrical effects, and is commonly referred to as the "pseudo-luminosity" \citep{arzoumanian2002,lorimer2006}. 

Following the flux calculation of every pulsar, we determine the beaming fraction of pulsars in a particular spin period bin.  \citet{tauris1998} defined an empirical relationship between a pulsar's spin period and the beaming fraction for pulsars where $P$\,>\,0.1\,s:

\begin{equation}
    \label{eqn:beaming_TM}
    f_\mathrm{TM}\text{(}P\text{)} \text{ = } 0.09\left[\log\left(\frac{P}{s}\right) - 1\right]^2\text{ + }0.03.
\end{equation}

\noindent Since a small portion of our simulated pulsars have $P$\,$\leq$\,0.1\,s we depict the beaming fraction for pulsars in this regime by a linear relationship.  The relationship is defined by a beaming fraction of 0.9 to 0.39 for 0.001\,$\leq$\,$P$(s)\,$\leq$\,0.1.  The lower limit was chosen from $f_{TM}$($0.1$)\,=\,0.39, while the upper limit of 0.9 was determined from millisecond pulsars by \citet{kramer1998}.  From these limits we extend the \citet{tauris1998} beaming fraction linearly with

\begin{equation}
    \label{eqn:beaming_Kramer}
    f_\mathrm{TMe}\text{(}P\text{)} \text{ = } -0.255\text{ }\log\left(\frac{P}{s}\right) \text{ + }0.135, 
\end{equation}

\noindent where  0.001\,$\leq$\,$P$(s)\,$\leq$\,0.1.  Finally, the total number of pulsars in a period bin is given by multiplying the number of pulsars with the beaming fraction.  Furthermore, to test how sensitive the number of detected pulsars is to the default empirical radio luminosity (Equation~\ref{eqn:lum}, $\sigma_{L_\mathrm{corr}}$\,=\,0.8) and the beaming fraction model (Equation~\ref{eqn:beaming_TM} \& \ref{eqn:beaming_Kramer}), we consider a second model in each case.  For the radio luminosity the second model is defined by adapting $\sigma_{L_\mathrm{corr}}$ from 0.8 to 2, simply to test how sensitive the number of detected pulsars is to $\sigma_{L_\mathrm{corr}}$.  The second empirical beaming fraction model was derived by \citet{lyne1988}: 

\begin{equation}
    \label{eqn:beaming_LM}
    f_\mathrm{LM}\text{(}P\text{)} \text{ = } \frac{4}{\pi}  \text{} \sin\left(6.5^\circ P^{-1\text{/}3}\right),
\end{equation}

\noindent and is valid for spin periods down to 0.005\,s.  For pulsars with $P$\,<\,0.005\,s  we set $f_\mathrm{LM}$(0.005)\,=\,0.78. 





\subsection{Star formation model}\label{sec:norm}

To account for the star formation history (SFH) of the SMC, we normalise our simulation with respect to the simulated mass and the mass available for each SFE ($M_\mathrm{SFE}$).  The mass available for star formation (SF) is calculated for two star formation histories (SFHs) determined by \citet{harris2004} and  \citet{antoniou2010}.   \citet{harris2004} composed an averaged photometric SFH (HZ04) of the SMC using the Magellanic Clouds Photometric Survey (MCPS, \citealt{zaritsky2002}, 2004).  Their data suggested that the most recent SFE occurred 60\,Myr ago with a star formation rate (SFR) of 0.28\,M$_\odot$\,yr\textsuperscript{-1} for 31\,Myr.  The star formation was averaged across time and spatial bins for the entire SMC.  The SFEs for the last 100\,Myr were not resolved, implying that there could be unresolved, intense star formation bursts.  \citet{harris2004} also calculated the metallicity evolution with every SFE, and found that the 60\,Myr SFE has a metallicity of Z\,=\,0.008. This is a factor of two larger than is usually assumed.  Later on \citet{antoniou2010} derived the SFH (A10) of regions in the SMC containing X-ray binaries from the spatially resolved HZ04 SFH.  The A10 SFH resolved the SFH of the last 100\,Myr into four star formation bursts.  As in the previous study they found evidence for a SFE, 67\,Myr ago, but with an average SFR of 0.72\,M$_\odot$\,yr\textsuperscript{-1}, followed by a SFE 42\,Myr ago with a SFR 1.72\,M$_\odot$\,yr\textsuperscript{-1}.  They concluded that the most recent SFEs occured 17\,Myr and 11\,Myr ago with SFRs of 0.38\,M$_\odot$\,yr\textsuperscript{-1} and 0.85\,M$_\odot$\,yr\textsuperscript{-1} respectively.  The mass available for star formation for a given SFH is calculated by multiplying the duration ($\Delta$t) of the SFEs with the corresponding SFRs (Table~\ref{tab:sfh}).



The normalisation also depends on the simulated mass of the primordial binaries ($M_{total}$):

\begin{equation}
    \label{eqn:mtotal}
    M_{total} \text{ = } \overbrace{\left[\frac{N_b}{f_{param}}\times M_b\right]}^{\text{Binaries}} \text{ + }  \overbrace{\left[\frac{N_b}{f_{param}}\times\left(\frac{1-f_{bin}}{f_{bin}}\right)\times M_p\right]}^{\text{Single stars}},
\end{equation}

\noindent where $N_b$ is the 3 million simulated binaries, while $M_b$\,=\,0.74\,M$_\odot$ and $M_p$\,=\,0.49\,M$_\odot$ represent the average, simulated binary and primary mass respectively, and $f_{bin}$ refers to the binary fraction.  The total simulation mass also accounts for the fraction of the total SMC stellar population we simulated ($f_\mathrm{param}$) by assuming the population consists of stars in a mass range from 0.1\,\textendash\,100\,M$_\odot$.  Since the primaries consist of stars with masses between 4.4 and 40\,M$_\odot$, only 0.74\%  of the SMC's stellar population was simulated, i.e  $f_\mathrm{param}$\,=\,0.0074.  The total simulated mass depends critically on the binary fraction.  As we assume a 100\% binary fraction the \textit{single star} term in Equation~\ref{eqn:mtotal} becomes zero.  The scaling factor $c$ is given by:

\begin{equation}
    c \text{ = } \frac{M_\mathrm{SFE}}{M_{total}}.
\end{equation}

\noindent  The total number of simulated pulsars normalised with respect to the mass available for star formation is determined by multiplying $c$ with the number of simulated pulsars.  Table~\ref{tab:sfh} lists the SFH parameters calculated by HZ04 and A10, as well as $c$ for every SFE.  For the default model we adopt the A10 SFH.


\begin{table}
\centering
\caption{SFH parameters of the SMC as reported by \citet{harris2004} and \citet{antoniou2010}, as well as the calculated SFE mass scaling factor $c$.}
\label{tab:sfh}
\begin{tabular}{llllll}
\hline
SF$^a$ Peak & SFE$^b$ & SFR & $\Delta$t $^c$ & $M_{SFR}$ & $c$\\
(Myr) & (Myr) & (M$_\odot$\,yr\textsuperscript{-1}) & (Myr) & (M$_\odot$) & (Number)\\
\hline
\hline
\multicolumn{6}{|c|}{\citet{harris2004}} \\
\hline
60 & 53\,--\,84 & 0.28 & 31 & 8.7\,$\times$10$^6$ & 0.029 \\
\hline
\multicolumn{6}{|c|}{\citet{antoniou2010}} \\
\hline
11 & 7\,--\,15 & 0.85 & 8 & 6.8\,$\times$10$^6$ & 0.023 \\
17 & 9.5\,--\,24.5 & 0.38 & 15 & 5.7\,$\times$10$^6$ & 0.019 \\
42 & 25\,--\,59 & 1.72 & 34 & 58.5\,$\times$10$^6$ & 0.196 \\
68 & 41\,--\,95 & 0.72 & 54 & 38.9\,$\times$10$^6$ & 0.130 \\
\hline
\end{tabular}
\begin{tablenotes}
\item$^a$SF: Star formation.
\item$^b$SFE: The onset and termination of each SFE.
\item$^c$$\Delta$t: SFE duration.
\end{tablenotes}
\end{table}

\subsection{Survey detection limits}

Besides the total number of pulsars currently present in the SMC, we also determine the number of synthetic pulsars that can be detected by past and future SMC surveys.  We model the specific selection effects of the \citet{manchester2006}, and a future SMC MeerKAT survey.  The sensitivity of a survey depends on the instrument used, as well as the observational setup.  The Manchester survey was conducted with the 21\,cm multi-beam receiver \citep{staveley1996} on the Parkes radio telescope.  The gain for the centre beam is 0.69\,K\,Jy\textsuperscript{-1}, and system temperature is 24\,K.  A potential MeerKAT survey configuration combines 40 dishes in the core of the array, resulting in a gain of 1.6\,K\,Jy\textsuperscript{-1} and system temperature of 17\,K.  The proposed SMC MeerKAT survey is assumed to observe an area equivalent to the Manchester survey.
Prior to the Manchester survey, \citet{crawford2001} carried out an equally sensitive survey.  However, the Manchester survey was more complete, since it covered a larger region of the SMC, and the pulsar searches were conducted across a larger dispersion measure range.  For these reasons we will discuss the simulation results with respect to the Manchester survey, and not the Crawford survey.  Table~\ref{tab:surveys} lists the observational parameters that were used to generate the survey sensitivity curves.  The sensitivity limits set the threshold for the number of pulsars that are detected by a particular survey, and are taken at the centre of each spin period bin.  

\begin{table}
\centering
\caption{Observational setup of SMC radio pulsar surveys, as well as the limiting flux densities at 1400\,MHz for $P$\,$\geq$\,50\,ms.}
\label{tab:surveys}
\begin{tabular}{llllll}
\hline
Survey &T$_\mathrm{int}$&$t_\mathrm{samp}$&$\Delta\nu$& N$_{\Delta\nu}$$^a$&S$_{1400}$\\
 &(s)&($\mu$s)&(MHz)& (Number)&(mJy)\\
\hline
\hline
\citet{crawford2001} & 8\,400 & 250 & 288 & 96 & 0.066 \\
\citet{manchester2006} & 8\,400 & 1\,000 & 288 & 96 & 0.067 \\
MeerKAT  & 7\,200 & 64 & 800 & 4096 & 0.012\\
\hline
\end{tabular}
\begin{tablenotes}
\item$^a$N$_{\Delta\nu}$:  Number of frequency channels.
\end{tablenotes}
\end{table}

\section{Results}



\subsection{Default model}

\begin{figure*}
    \centering
    \includegraphics[width=.75\textwidth]{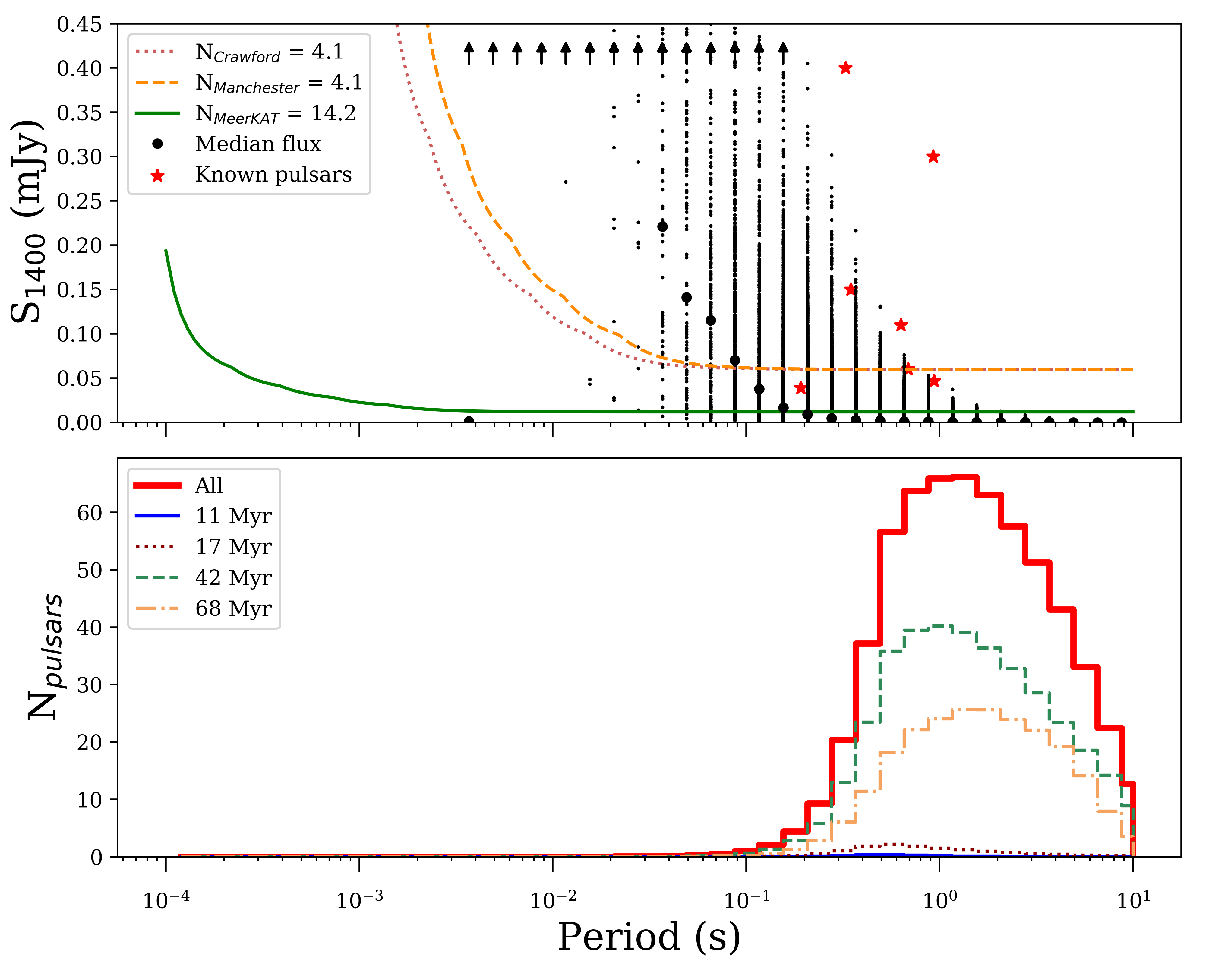}
    \caption{Default model of pulsars in binaries, adopting the \citet{tauris1998} beaming fraction and, $\sigma_{L_{corr}}$\,=\,0.8 luminosity model.  The top panel exhibits the various survey sensitivity curves with the corresponding number of detected synthetic pulsars in the legend, as well as the brightness distribution of the pulsars.  The small black points represent the brightness of individual pulsars, while the large black dots show the median flux of pulsars in a period bin.  The black arrows indicate points that are off the plot.  The red stars denote the known pulsar population.  The bottom panel shows the period distribution of all pulsars in binaries in the SMC that formed during the respective SFEs.}
    \label{fig:sensitivity_detached}
\end{figure*}
\begin{figure*}
    \centering
    \includegraphics[width=.75\textwidth]{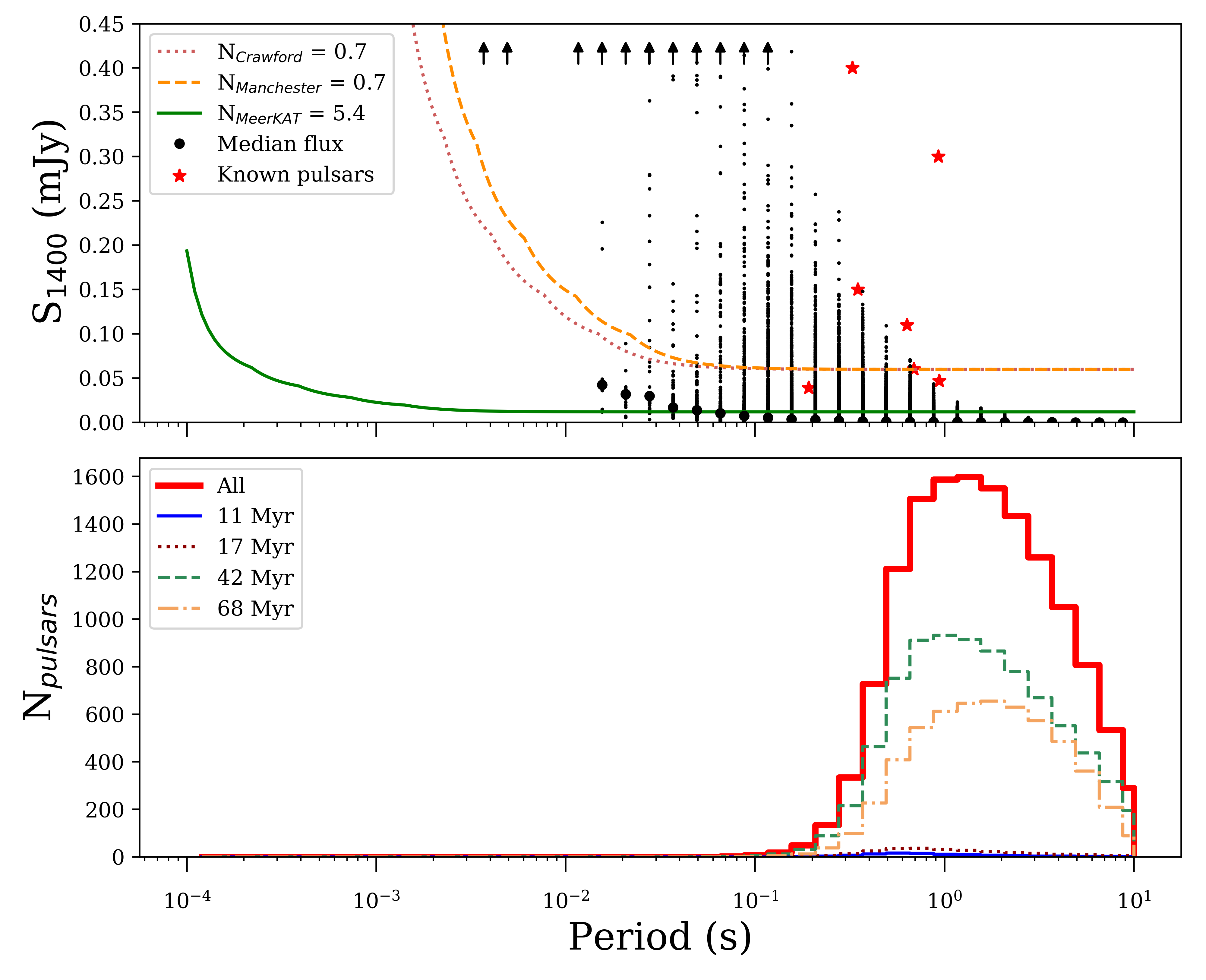}
    \caption{Default model of isolated pulsars, adopting the \citet{tauris1998} beaming fraction and, $\sigma_{L_{corr}}$\,=\,0.8 luminosity model.  The top panel exhibits the various survey sensitivity curves with the corresponding number of detected synthetic pulsars in the legend, as well as the brightness distribution of the pulsars.  The small black points represent the brightness of individual pulsars, while the large black dots show the median flux of pulsars in a period bin.  The black arrows indicate points that are off the plot.  The red stars denote the known pulsar population.  The bottom panel shows the period distribution of all isolated pulsars in the SMC that formed during the respective SFEs.}
    \label{fig:sensitivity_disrupted}
\end{figure*}

When assuming the default model (see Section~\ref{sec:initial_conditions}) the SMC is predicted to have formed 1.5\,$\times$10\textsuperscript{4} observable\footnote{Pulsars beaming towards us.} pulsars, and 1.8\,$\times$10\textsuperscript{5} pulsars in total.  Of these pulsars we predict that the Manchester survey will detect 4.8 synthesised pulsars, while the future MeerKAT survey is predicted to detect 19.6 pulsars.  Figure~\ref{fig:sensitivity_detached} and \ref{fig:sensitivity_disrupted} compares the brightness and period distributions of pulsars in binaries, as well as isolated pulsars.  These figures illustrate the low radio flux of most synthesised pulsars, rendering them undetectable by the previous survey limits, as well as the MeerKAT survey. However, the spin period distribution of pulsars peaks at $\sim$1s. So, it is the combined impact of the prevalence of pulsars around the 1s spin period together with the increased brightness of the faster spinning pulsars (i.e. >\,0.1s) that explains the observed period and luminosity distribution of the pulsars detected by the Manchester survey. (See Section~\ref{sec:4.1} for further detail.)

\subsection{Sensitivity to initial conditions}\label{sec:results_additional}

We carried out 96 simulations for two SFH to determine the effect the various parameters have on the total number of synthetic pulsars, as well as the number of pulsars recovered by the Manchester and MeerKAT surveys.  The assumed SFH and radio luminosity model are the main parameters that have a significant effect on the synthesised radio pulsar population of the SMC.  

 The models where the A10 SFH was assumed with the empirical luminosity model ($\sigma_{\text{Lcorr}}$\,=\,0.8), while varying the CE model, CE parameter, metallicity, and beaming fraction resulted in a range of synthesised pulsars (see Appendix~\ref{sec:tables}).  A minimum of 1.2\,$\times$10\textsuperscript{4} and a maximum of 2.2\,$\times$10\textsuperscript{4} pulsars are synthesised.  Of these synthesised pulsars, the Manchester survey is predicted to detect 2.8\,\textendash\,6.0 pulsars, while the future MeerKAT survey is predicted to detect 12.3\,\textendash\,23.0 pulsars. 

\subsubsection{Star formation history}

The number of synthesised and detected pulsars are exceptionally sensitive to the assumed SFH.  The default A10 SFH produces $\sim$12 times more pulsars when compared to the HZ04 SFH.  This is directly related to the mass available for star formation. The A10 SFH provides 110\,$\times$\,10\textsuperscript{6}\,M$_\odot$ for SF, while the HZ04 SFH results in 8.7\,$\times$\,10\textsuperscript{6}\,M$_\odot$.  Consequently, the A10 SFH has 12.6 times more mass available for star formation than the HZ04 SFH, and in turn $\sim$12 times more pulsars are synthesised.  The assumed SFH also has consequences for the number of recovered pulsars.  The A10 SFH results in the recovery of 14.5 and 59.4 times more synthesised pulsars for the Manchester and MeerKAT survey when compared to the HZ04 SFH.  The HZ04 SFH yields an insignificant population of pulsars ($\sim$1$\,\times$\,10\textsuperscript{3}), thus we will only report the results of the A10 SFH.  The results of the 96 simulations for the A10 SFH is summarised in Appendix~\ref{sec:tables}.  

\subsubsection{Radio luminosity model}

The number of recovered pulsars is contingent on the assumed radio luminosity model.  The alternate luminosity model ($\sigma_{L_{corr}}$\,=\,2.0) allows for brighter synthesised pulsars when compared to the empirical, default model ($\sigma_{L_\mathrm{corr}}$\,=\,0.8).  Consequently, more pulsars are recovered when the survey selection effects are applied with the alternate luminosity model.  In particular the alternate radio luminosity model results in the detection of 16.0 synthesised pulsars with the Manchester survey, while the MeerKAT survey is predicted to detect 112.6 pulsars.  This corresponds to an increase by a factor of 3.3 and 5.7 for the Manchester and MeerKAT surveys respectively, when adopting the alternate luminosity model instead of the empirical, default model's radio luminosity.


\subsection{Consistent population properties}

The simulations are extremely robust in terms of the CE model ($\gamma\alpha$ vs $\alpha\alpha$), CE parameter, metallicity, and beaming fraction, i.e across all these models there are no categorical differences in the total number of synthesised or recovered pulsars.  However, the applied SFH and radio luminosity model have a significant impact on both the number of synthesised and recovered pulsars. As such, we group the results by the A10 SFH as well as the empirical radio luminosity model ($\sigma_{L_\mathrm{corr}}$\,=\,0.8), and determine the various medians with corresponding standard deviations as error estimates.  

On average the SMC synthesised (1.6\,$\pm$\,0.3)\,$\times$10\textsuperscript{4} detectable radio pulsars in the last $\sim$100\,Myr.  This population excludes old, recycled MSPs, and only includes pulsars beaming towards us.  By applying the Manchester survey detection limits we recover 4.0\,$\pm$\,0.8 synthetic pulsars on average, while the SMC MeerKAT survey is predicted to detect 17.2\,$\pm$\,2.5 pulsars.

\subsection{Pulsars in binaries}

The NS kick velocity distribution does not have a significant impact on the number of synthesised or recovered pulsars.  However, it predicts different fractions of synthesised pulsars in binaries.  When applying NS kick velocity distributions from \citet{hobbs2005} and \citet{verbunt2017} an average of 2.4\% and 5.5\% of the pulsar population is predicted to occur in binaries, respectively.  Conversely, when adopting the alternative model where we reduced the kick velocities from \citet{verbunt2017} by a factor of 10 an average of 19\% of the synthesised pulsar population is retained within a binary.  The pulsars in binaries with high mass companions (M\,>\,5\,M$_\odot$) are primordial high mass X-ray binaries (HMXBs).  When considering the default model 4\% of pulsars are predicted to be in binaries, implying $\sim$160 synthesised pulsars are in binaries with high mass companions with orbital periods ($P_{\text{orb}}$) below a 1000\,days.  Alternatively when applying the \citet{hobbs2005} distribution, the pulsar binary fraction is 2\% predicting $\sim$60 high mass binary pulsars. 

Although a relatively small number of synthesised pulsars occur in binaries, the simulations predict that $\sim$80\% of the synthesised pulsars that are detected by the Manchester survey are in binaries.  This prediction is inherent to the synthesised population of bright binaries with spin periods below 0.1\,s (Figure~\ref{fig:sensitivity_detached}).  These pulsars are young and typically short-lived, before the pulsar mechanism is quenched by binary interactions.  Pulsars within very compact binaries, and/or with massive companions are quenched shortly after formation when mass transfer ensues.  These pulsars have less time to spin down, and consequently are brighter on average than the isolated pulsars with spin periods below 0.1\,s. 

\begin{figure}
    \centering
    \includegraphics[width=\columnwidth]{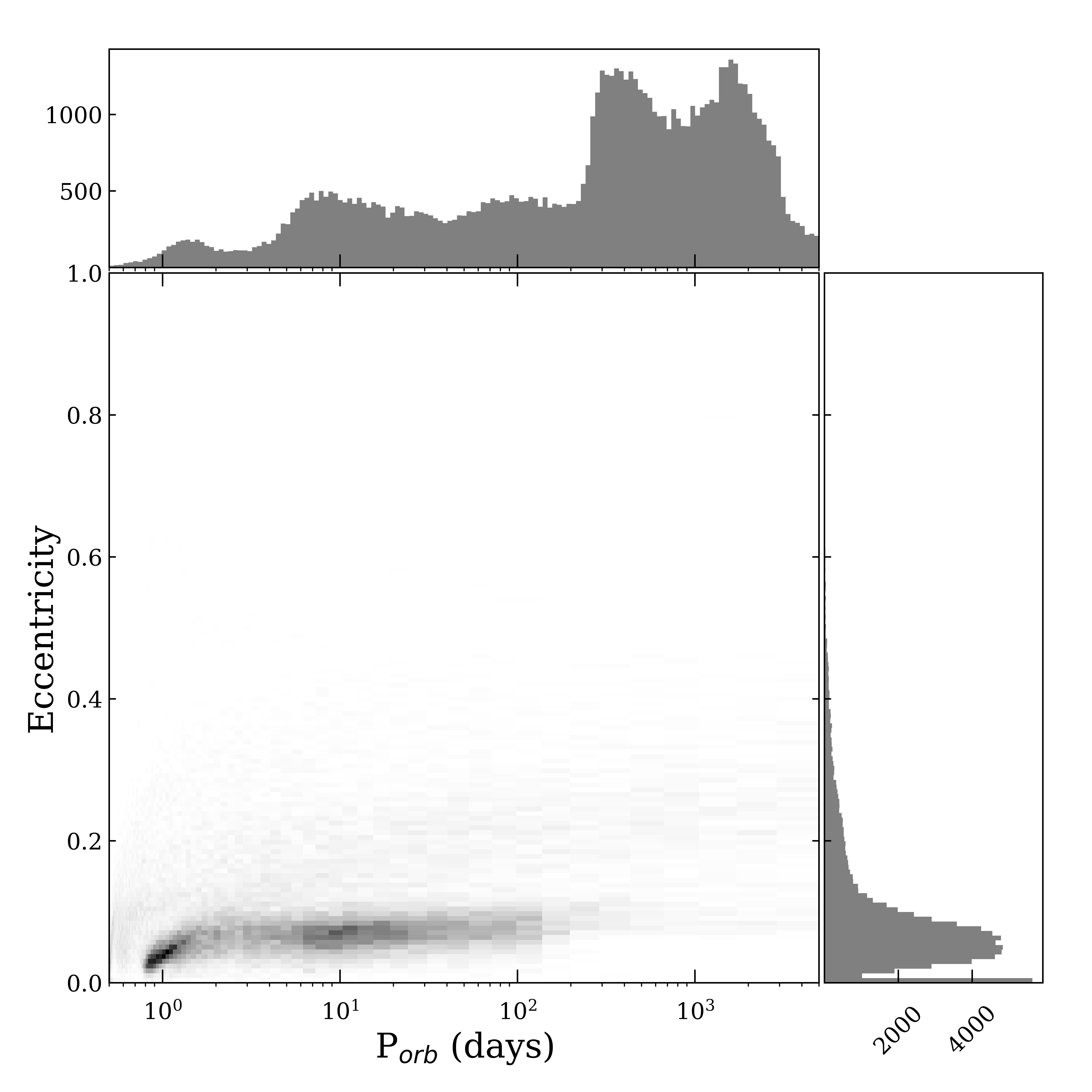}
    \caption{Intensity plot showing the relationship between the orbital period ($P_{\text{orb}}$) and eccentricity of pulsars in binaries.  The individual orbital period and eccentricity distributions are also depicted on the relevant axes.}
    \label{fig:porb_e}
\end{figure}

The orbital period ($P_{\text{orb}}$) and eccentricity ($e$) of the pulsars in binaries have a weak positive correlation (Figure~\ref{fig:porb_e}).  The binaries shown in the figure include 95\% of the binary population that have $P_{\text{orb}}$\,$\leq$\,5\,000\,days.  The orbital period distribution has five significant peaks, the peaks below $P_{\text{orb}}$\,$\sim$\,300\,days are associated with the SFEs of 11 \& 17\,Myr ago, while the remaining two peaks ($P_{\text{orb}}$\,$\sim$\,400\,\& 1200\,days) present the evolved binaries from the SFEs that occurred 42 \& 68\,Myr ago.  The evolved binary pulsars occur in wider binaries where minimal mass transfer occurs, extending the life time of the radio pulsars. There is a large number of binaries with a nearly circular orbit, after which the eccentricity distribution follows a Poisson-like distribution.  The peak at $e$\,$\sim$\,0 is associated with more evolved binaries from the 42 and 68\,Myr SFEs.  The most recent SFEs (11 \& 17\,Myr ago) have exponential eccentricity distributions with $\sim$18\% of their population having e\,>\,0.2, while the older SFEs (42 \& 68\,Myr ago) have fewer eccentric orbits with only $\sim$7\% of systems exceeding an eccentricity of 0.2.  

\section{Discussion}

\subsection{SMC radio pulsar population}\label{sec:4.1}

In this study we utilise binary population synthesis techniques to estimate the size of the SMC radio pulsar population, and apply various survey selection effects in an attempt to recover the observed population, and characterise the efficiency of the future SMC MeerKAT survey.  We predict that (1.6\,$\pm$\,0.3)\,$\times$10\textsuperscript{4} normal radio pulsars were formed in the SMC during the last $\sim$100\,Myr.  This population excludes recycled, old MSPs, and only includes pulsars beaming towards us.  This is in agreement with \citet{ridley2010}, who predict a population of (1.09\,$\pm$\,0.16)\,$\times$10\textsuperscript{4} normal pulsars in the SMC.  This agreement is striking, since our method was based on binary evolution and dependent on the SFH of the SMC, while \citet{ridley2010} utilised a Monte Carlo method \citep{lorimer2006} that seeded pulsars until the Manchester selection effects recovered five observed pulsars.  

\begin{figure}
    \centering
    \includegraphics[width=\columnwidth]{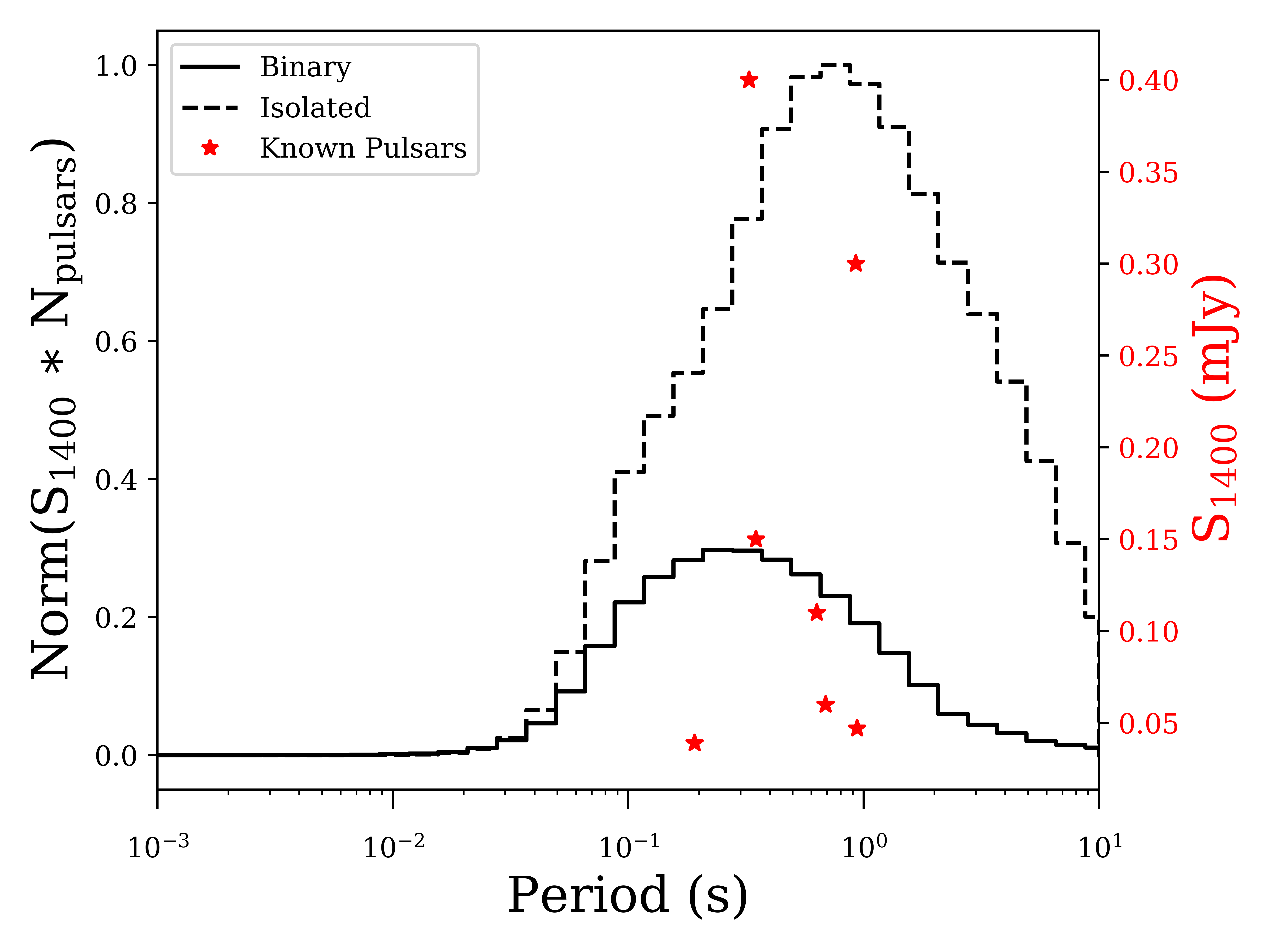}
    \caption{The normalised, bias corrected spin period distribution of pulsars in binaries and isolated pulsars (left y-scale).  This figure is a result of convolving the flux and period distributions, i.e. the top and bottom panels of Figure~\ref{fig:sensitivity_detached} and \ref{fig:sensitivity_disrupted}.  The stars show the flux of the known SMC pulsars (right y-scale).}
    \label{fig:convolve}
\end{figure}

We study how many pulsars may be observed by the Manchester survey, by applying the survey's specific selection effects and recover 4.0\,$\pm$\,0.8 synthetic pulsars, which compares well to the five pulsars observed with the \citet{manchester2006} survey.  The proposed SMC MeerKAT survey is $\sim$6 times more sensitive than the Manchester survey, however it is predicted to detect only 17.2\,$\pm$\,2.5 pulsars.  The large distance to the SMC reduces the flux of the radio pulsars to such an extent that MeerKAT will only detect the brightest pulsars.  To make a comparison between the observed and synthesised spin period distribution we convolve the simulated flux and spin period distributions (top and bottom panels of Figures~\ref{fig:sensitivity_detached} and \ref{fig:sensitivity_disrupted}) to create a normalised, bias corrected spin period distribution (Figure~\ref{fig:convolve}).  Currently there are only seven known pulsars in the SMC.  Such a small number of pulsars is insufficient to make extensive comparisons, however Figure~\ref{fig:convolve} shows that the observed pulsars are coincident with the peaks of the binary and isolated pulsar period distributions.  The majority of the isolated pulsars have a spin period distribution that peaks around 1\,s, and extends to 10\,s.  The pulsars in binaries account for only a few percent of all the simulated pulsars, and have a spin period distribution peaking at $\sim$0.3\,s.  Unlike the isolated pulsars, very few pulsars in binaries have spin periods beyond $\sim$1\,s.  Initially, rapidly spinning pulsars are prevented from accreting material by the propeller mechanism, however once they have spun down sufficiently the NS can accrete material and enter the HMXB phase.  The SMC in particular harbours a number of accreting NSs with spin periods below 10\,s (see \citealt{haberl2016}, and references therein).  Moreover, the simulations predict a population of $\sim$60\,\textendash\,160 NSs in binaries with massive companions.  These systems are primordial HMXBs, and compare well with the known population of $\sim$120 HMXBs in the SMC.

Another notable difference is that pulsars in binaries are brighter than isolated pulsars with spin periods below $\sim$0.1\,s.  As a result the simulations suggest that less sensitive surveys are more likely to detect pulsars in binaries, and that as many as 80\% of the Manchester pulsars are predicted to be in binaries.  Thus far one of the five pulsars has been identified as a pulsar in a binary with a B star in a 51\,day orbit \citep{kaspi1994}.  Due to the distance to the SMC extensive timing studies of the known pulsars have not been possible, thus the possibility that they are in wide orbits cannot be ruled out.  With the upcoming MeerKAT survey, timing studies are likely to identify new pulsars in binaries.  Especially since the bulk of the synthetic pulsars in binaries have orbital periods below $\sim$10\textsuperscript{3}\,days with fairly low eccentricities (Figure~\ref{fig:porb_e}) for which realistic observing campaigns can be launched to search for the binaries.  Apart from timing studies it is also possible to search for pulsars in binaries with acceleration searches.  \citet{ridley2013} carried out acceleration searches that were sensitive to pulsars in binaries with spin periods greater than 23.5\,ms on the archival \citet{manchester2006} data, but did not discovery any new pulsars in binaries. However, acceleration searches are most effective for close binaries, i.e. those more likely for MSPs, and not wide binaries.

\subsection{Constraints on assumed models}

In the following sections we place constraints on the SMC SFH, and the empirically determined radio luminosity model.

\subsubsection{Star formation history}

The A10 SFH successfully reproduced the predicted number of normal pulsars in the SMC, as well as the observed pulsar population.  This is in contrast with the HZ04 SFH, that neither produces the predicted number of pulsars nor recovers the observed pulsar population.  In fact the HZ04 SFH produces a virtually undetectable pulsar population.  The HZ04 is a global, spatially resolved, time averaged SFH, and hence does not account for short bursts of SF. It also does not resolve the SFH of the last 100\,Myr, which is when the majority of the pulsars would have formed.  The HZ04 may be more suitably applied to older stellar populations where a high temporal resolution is not required.  For our pulsar population it is essential to apply a SFH that is resolved throughout the last $\sim$100\,Myr, making the A10 SFH more appropriate.  The A10 SFH is also limited and not ideal, since it was not calculated for the entire SMC, but for regions containing X-ray binaries, which we then assumed for the entire SMC.  Regardless of which SFH is more appropriate, the simulations require a SFH that provides sufficient mass to synthesise enough pulsars for the Manchester survey to recover the five observed pulsars.  Thus, we require a SFH that provides at least as much mass as the modified A10 SFH (110\,$\times$\,10\textsuperscript{6}\,M$_\odot$). 

\subsubsection{Radio luminosity model}

The default model which adopts the empirical radio luminosity model ($\sigma_{L_\mathrm{corr}}$\,=\,0.8) produces sufficient pulsars to recover the five observed pulsars with a $\sim$20\% error when the Manchester selection effects are applied.  Reproducing the observed population with such accuracy provides confidence for the method, its results, and the predicted number of pulsars MeerKAT will detect.  The radio luminosity dispersion of the Galactic pulsars is well modelled for $\sigma_{L_\mathrm{corr}}$\,=\,0.8.  There is no observational evidence that the SMC pulsars will deviate from this relation.  Moreover, when adopting the alternate radio luminosity model ($\sigma_{L_\mathrm{corr}}$\,=\,2), it does not result in the recovery of the five observed pulsars, instead it overestimates the population by a factor of three.  The alternate model is not supported by observations, it was a theoretical test to determine how sensitive the simulations are to the assumed luminosity model.  Consequently, there are presently no clear indication that the SMC pulsar radio luminosities differ when compared to the Galactic pulsar population luminosities.

\subsubsection{Spontaneous magnetic field decay}

The same spontaneous magnetic field decay model was applied to all the simulations.  We modelled the spontaneous magnetic field decay of the radio pulsars by applying an Ohmic dissipation recipe with a decay time scale of a 100\,Myr, and predict that the Manchester survey could have detected $\sim$5 pulsars, which is in agreement with the 5 pulsars the SMC Manchester survey observed.  By including spontaneous magnetic field decay pulsars evolve faster and cross the death line sooner, i.e. by applying field decay the average life time of the pulsars are reduced.  If field decay is not important our results will underestimate the observable radio pulsar population of the SMC, and consequently also underestimate the predicted number of pulsars the Manchester survey could have detected, as well as the number of pulsars MeerKAT will detect.  Conversely, if the magnetic field decays faster ($\tau$\,<\,100\,Myr) we would have overestimated the number of observable and detected pulsars by the Manchester and MeerKAT survey.  However, given the agreement between our simulations and observations it is unlikely that spontaneous magnetic field decay occurs over a shorter time scale.  Due to the small observational sample, and the long decay time scale we cannot firmly conclude if magnetic field decay occurs at all.

\section{Conclusion}

We modelled the present day radio pulsar population of the SMC (excluding millisecond pulsars).  On average the number of synthesised and recovered pulsars were not affected by metallicity, CE evolution, NS kick velocity distribution or the beaming fraction model, but are strongly dependent on the recent star formation rate and the pulsar luminosity model. The simulations predict a population of (1.6\,$\pm$\,0.3)\,$\times$10\textsuperscript{4} normal radio pulsars.  When applying the selection effects of the \citet{manchester2006} survey we recover 4.0\,$\pm$\,0.8 synthetic pulsars, which is in agreement with the five observed pulsars.  Furthermore, the future SMC MeerKAT survey is predicted to detect 17.2\,$\pm$\,2.5 pulsars.  This implies that the proposed SMC MeerKAT survey will not be sensitive enough to detect a significant portion of the SMC pulsar population.  That being said MeerKAT will provide a valuable test to the predicted number of detectable pulsars, and contribute to our understanding of massive binary evolution.



\section*{Acknowledgements}
NT, VMcB, and DAHB acknowledges support of the National Research Foundation of South Africa (grants 98969, 93405, and 96094).  ST acknowledges support from the Netherlands Research Council NWO (grant VENI [nr. 639.041.645]).  BWS acknowledge funding from the European Research Council (ERC) under the European Union's Horizon 2020 research and innovation programme (grant agreement No. 694745).



\bibliographystyle{mn2e}
\bibliography{ref.bib}

\begin{thebibliography}{}

\bibitem[\protect\citeauthoryear{{Abt}}{{Abt}}{1983}]{abt1983}
{Abt} H.~A.,  1983, \araa, 21, 343

\bibitem[\protect\citeauthoryear{{Antoniou}, {Zezas}, {Hatzidimitriou} \&
  {Kalogera}}{{Antoniou} et~al.}{2010}]{antoniou2010}
{Antoniou} V.,  {Zezas} A.,  {Hatzidimitriou} D.,    {Kalogera} V.,  2010,
  \apjl, 716, L140

\bibitem[\protect\citeauthoryear{{Arzoumanian}, {Chernoff} \&
  {Cordes}}{{Arzoumanian} et~al.}{2002}]{arzoumanian2002}
{Arzoumanian} Z.,  {Chernoff} D.~F.,    {Cordes} J.~M.,  2002, \apj, 568, 289

\bibitem[\protect\citeauthoryear{{Bates}, {Lorimer}, {Rane} \&
  {Swiggum}}{{Bates} et~al.}{2014}]{bates2014}
{Bates} S.~D.,  {Lorimer} D.~R.,  {Rane} A.,    {Swiggum} J.,  2014, \mnras,
  439, 2893

\bibitem[\protect\citeauthoryear{{Bhattacharya}, {Wijers}, {Hartman} \&
  {Verbunt}}{{Bhattacharya} et~al.}{1992}]{bhattacharya1992}
{Bhattacharya} D.,  {Wijers} R. A.~M.~J.,  {Hartman} J.~W.,    {Verbunt} F.,
  1992, \aap, 254, 198

\bibitem[\protect\citeauthoryear{{Blaauw}}{{Blaauw}}{1961}]{blaauw1961}
{Blaauw} A.,  1961, Bull. Astron. Inst. Netherlands, 15, 265

\bibitem[\protect\citeauthoryear{{Bray} \& {Eldridge}}{{Bray} \&
  {Eldridge}}{2016}]{bray2016}
{Bray} J.~C.,  {Eldridge} J.~J.,  2016, \mnras, 461, 3747

\bibitem[\protect\citeauthoryear{{Camacho}, {Torres}, {Garc{\'\i}a-Berro},
  {Zorotovic}, {Schreiber}, {Rebassa-Mansergas}, {Nebot G{\'o}mez-Mor{\'a}n} \&
  {G{\"a}nsicke}}{{Camacho} et~al.}{2014}]{camacho2014}
{Camacho} J.,  {Torres} S.,  {Garc{\'\i}a-Berro} E.,  {Zorotovic} M.,
  {Schreiber} M.~R.,  {Rebassa-Mansergas} A.,  {Nebot G{\'o}mez-Mor{\'a}n} A.,
    {G{\"a}nsicke} B.~T.,  2014, \aap, 566, A86

\bibitem[\protect\citeauthoryear{{Chen} \& {Ruderman}}{{Chen} \&
  {Ruderman}}{1993}]{chen1993}
{Chen} K.,  {Ruderman} M.,  1993, \apj, 402, 264

\bibitem[\protect\citeauthoryear{{Cie{\'s}lar}, {Bulik} \&
  {Os{\l}owski}}{{Cie{\'s}lar} et~al.}{2020}]{cieslar2020}
{Cie{\'s}lar} M.,  {Bulik} T.,    {Os{\l}owski} S.,  2020, \mnras, 492, 4043

\bibitem[\protect\citeauthoryear{{Cordes}, {Romani} \& {Lundgren}}{{Cordes}
  et~al.}{1993}]{cordes1993}
{Cordes} J.~M.,  {Romani} R.~W.,    {Lundgren} S.~C.,  1993, \nat, 362, 133

\bibitem[\protect\citeauthoryear{{Crawford}, {Kaspi}, {Manchester}, {Lyne},
  {Camilo} \& {D'Amico}}{{Crawford} et~al.}{2001}]{crawford2001}
{Crawford} F.,  {Kaspi} V.~M.,  {Manchester} R.~N.,  {Lyne} A.~G.,  {Camilo}
  F.,    {D'Amico} N.,  2001, \apj, 553, 367

\bibitem[\protect\citeauthoryear{{Crowl}, {Sarajedini}, {Piatti}, {Geisler},
  {Bica}, {Clari{\'a}} \& {Santos} Jo{\~a}o F.~C.}{{Crowl}
  et~al.}{2001}]{crowl2001}
{Crowl} H.~H.,  {Sarajedini} A.,  {Piatti} A.~E.,  {Geisler} D.,  {Bica} E.,
  {Clari{\'a}} J.~J.,    {Santos} Jo{\~a}o F.~C. J.,  2001, \aj, 122, 220

\bibitem[\protect\citeauthoryear{{Cumming}, {Arras} \& {Zweibel}}{{Cumming}
  et~al.}{2004}]{cumming2004}
{Cumming} A.,  {Arras} P.,    {Zweibel} E.,  2004, \apj, 609, 999

\bibitem[\protect\citeauthoryear{{de Kool}}{{de Kool}}{1990}]{dekool1990}
{de Kool} M.,  1990, \apj, 358, 189

\bibitem[\protect\citeauthoryear{{de Kool}, {van den Heuvel} \& {Pylyser}}{{de
  Kool} et~al.}{1987}]{dekool1987}
{de Kool} M.,  {van den Heuvel} E.~P.~J.,    {Pylyser} E.,  1987, \aap, 183, 47

\bibitem[\protect\citeauthoryear{{Duch{\^e}ne} \& {Kraus}}{{Duch{\^e}ne} \&
  {Kraus}}{2013}]{duchene2013}
{Duch{\^e}ne} G.,  {Kraus} A.,  2013, \araa, 51, 269

\bibitem[\protect\citeauthoryear{{Faucher-Gigu{\`e}re} \&
  {Kaspi}}{{Faucher-Gigu{\`e}re} \& {Kaspi}}{2006}]{faucher2006}
{Faucher-Gigu{\`e}re} C.-A.,  {Kaspi} V.~M.,  2006, \apj, 643, 332

\bibitem[\protect\citeauthoryear{{Geppert}}{{Geppert}}{2009}]{geppert2009}
{Geppert} U.,  2009, {Turning Points in the Evolution of Isolated Neutron
  Stars'Magnetic Fields}.
p.~319

\bibitem[\protect\citeauthoryear{{Gonthier}, {Van Guilder} \&
  {Harding}}{{Gonthier} et~al.}{2004}]{gonthier2004}
{Gonthier} P.~L.,  {Van Guilder} R.,    {Harding} A.~K.,  2004, \apj, 604, 775

\bibitem[\protect\citeauthoryear{{Graczyk} et~al.,}{{Graczyk}
  et~al.}{2014}]{graczyk2014}
{Graczyk} D.  et~al., 2014, \apj, 780, 59

\bibitem[\protect\citeauthoryear{{Gull{\'o}n}, {Miralles}, {Vigan{\`o}} \&
  {Pons}}{{Gull{\'o}n} et~al.}{2014}]{gullon2014}
{Gull{\'o}n} M.,  {Miralles} J.~A.,  {Vigan{\`o}} D.,    {Pons} J.~A.,  2014,
  \mnras, 443, 1891

\bibitem[\protect\citeauthoryear{{Gull{\'o}n}, {Pons}, {Miralles},
  {Vigan{\`o}}, {Rea} \& {Perna}}{{Gull{\'o}n} et~al.}{2015}]{gullon2015}
{Gull{\'o}n} M.,  {Pons} J.~A.,  {Miralles} J.~A.,  {Vigan{\`o}} D.,  {Rea} N.,
     {Perna} R.,  2015, \mnras, 454, 615

\bibitem[\protect\citeauthoryear{{Gunn} \& {Ostriker}}{{Gunn} \&
  {Ostriker}}{1970}]{gunn1970}
{Gunn} J.~E.,  {Ostriker} J.~P.,  1970, \apj, 160, 979

\bibitem[\protect\citeauthoryear{{Gurevich} \& {Istomin}}{{Gurevich} \&
  {Istomin}}{2007}]{gurevich2007}
{Gurevich} A.~V.,  {Istomin} Y.~N.,  2007, \mnras, 377, 1663

\bibitem[\protect\citeauthoryear{{Haberl} \& {Pietsch}}{{Haberl} \&
  {Pietsch}}{2004}]{haberl2004}
{Haberl} F.,  {Pietsch} W.,  2004, \aap, 414, 667

\bibitem[\protect\citeauthoryear{{Haberl} \& {Sturm}}{{Haberl} \&
  {Sturm}}{2016}]{haberl2016}
{Haberl} F.,  {Sturm} R.,  2016, \aap, 586, A81

\bibitem[\protect\citeauthoryear{{Harris} \& {Zaritsky}}{{Harris} \&
  {Zaritsky}}{2004}]{harris2004}
{Harris} J.,  {Zaritsky} D.,  2004, \aj, 127, 1531

\bibitem[\protect\citeauthoryear{{Heger}, {Fryer}, {Woosley}, {Langer} \&
  {Hartmann}}{{Heger} et~al.}{2003}]{heger2003}
{Heger} A.,  {Fryer} C.~L.,  {Woosley} S.~E.,  {Langer} N.,    {Hartmann}
  D.~H.,  2003, \apj, 591, 288

\bibitem[\protect\citeauthoryear{{Heggie}}{{Heggie}}{1975}]{heggie1975}
{Heggie} D.~C.,  1975, \mnras, 173, 729

\bibitem[\protect\citeauthoryear{{Hobbs}, {Lorimer}, {Lyne} \&
  {Kramer}}{{Hobbs} et~al.}{2005}]{hobbs2005}
{Hobbs} G.,  {Lorimer} D.~R.,  {Lyne} A.~G.,    {Kramer} M.,  2005, \mnras,
  360, 974

\bibitem[\protect\citeauthoryear{{Huang}}{{Huang}}{1963}]{huang1963}
{Huang} S.-S.,  1963, \apj, 138, 471

\bibitem[\protect\citeauthoryear{{Igoshev} \& {Popov}}{{Igoshev} \&
  {Popov}}{2015}]{igoshev2015}
{Igoshev} A.~P.,  {Popov} S.~B.,  2015, Astronomische Nachrichten, 336, 831

\bibitem[\protect\citeauthoryear{{Ivanova} et~al.,}{{Ivanova}
  et~al.}{2013}]{ivanova2013}
{Ivanova} N.  et~al., 2013, A\&A Rev., 21, 59

\bibitem[\protect\citeauthoryear{{Janka}}{{Janka}}{2012}]{janka2012}
{Janka} H.-T.,  2012, Annual Review of Nuclear and Particle Science, 62, 407

\bibitem[\protect\citeauthoryear{{Johnston} \& {Karastergiou}}{{Johnston} \&
  {Karastergiou}}{2017}]{johnston2017}
{Johnston} S.,  {Karastergiou} A.,  2017, \mnras, 467, 3493

\bibitem[\protect\citeauthoryear{Kaspi, Johnston, Bell, Manchester, Bailes,
  Bessell, Lyne \& D'Amico}{Kaspi et~al.}{1994}]{kaspi1994}
Kaspi V.~M.,  Johnston S.,  Bell J.~F.,  Manchester R.~N.,  Bailes M.,  Bessell
  M.,  Lyne A.~G.,    D'Amico N.,  1994, ApJ, 423, 44

\bibitem[\protect\citeauthoryear{{Knigge}, {Coe} \& {Podsiadlowski}}{{Knigge}
  et~al.}{2011}]{knigge2011}
{Knigge} C.,  {Coe} M.~J.,    {Podsiadlowski} P.,  2011, \nat, 479, 372

\bibitem[\protect\citeauthoryear{{Kobulnicky} \& {Fryer}}{{Kobulnicky} \&
  {Fryer}}{2007}]{kobulnicky2007}
{Kobulnicky} H.~A.,  {Fryer} C.~L.,  2007, \apj, 670, 747

\bibitem[\protect\citeauthoryear{{Kouwenhoven}, {Brown}, {Portegies Zwart} \&
  {Kaper}}{{Kouwenhoven} et~al.}{2007}]{kouwenhoven2007}
{Kouwenhoven} M.~B.~N.,  {Brown} A.~G.~A.,  {Portegies Zwart} S.~F.,    {Kaper}
  L.,  2007, \aap, 474, 77

\bibitem[\protect\citeauthoryear{{Kramer}, {Xilouris}, {Lorimer}, {Doroshenko},
  {Jessner}, {Wielebinski}, {Wolszczan} \& {Camilo}}{{Kramer}
  et~al.}{1998}]{kramer1998}
{Kramer} M.,  {Xilouris} K.~M.,  {Lorimer} D.~R.,  {Doroshenko} O.,  {Jessner}
  A.,  {Wielebinski} R.,  {Wolszczan} A.,    {Camilo} F.,  1998, \apj, 501, 270

\bibitem[\protect\citeauthoryear{{Kroupa}}{{Kroupa}}{2001}]{kroupa2001}
{Kroupa} P.,  2001, \mnras, 322, 231

\bibitem[\protect\citeauthoryear{{Lai}}{{Lai}}{2001}]{lai2001}
{Lai} D.,  2001, {Neutron Star Kicks and Asymmetric Supernovae}.
p.~424

\bibitem[\protect\citeauthoryear{{Leonard}, {Hills} \& {Dewey}}{{Leonard}
  et~al.}{1994}]{leonard1994}
{Leonard} P. J.~T.,  {Hills} J.~G.,    {Dewey} R.~J.,  1994, \apjl, 423, L19

\bibitem[\protect\citeauthoryear{{Linden}, {Sepinsky}, {Kalogera} \&
  {Belczynski}}{{Linden} et~al.}{2009}]{linden2009}
{Linden} T.,  {Sepinsky} J.~F.,  {Kalogera} V.,    {Belczynski} K.,  2009,
  \apj, 699, 1573

\bibitem[\protect\citeauthoryear{{Livio} \& {Soker}}{{Livio} \&
  {Soker}}{1988}]{livio1988}
{Livio} M.,  {Soker} N.,  1988, \apj, 329, 764

\bibitem[\protect\citeauthoryear{{Lorimer}, {Bailes}, {Dewey} \&
  {Harrison}}{{Lorimer} et~al.}{1993}]{lorimer1993}
{Lorimer} D.~R.,  {Bailes} M.,  {Dewey} R.~J.,    {Harrison} P.~A.,  1993,
  \mnras, 263, 403

\bibitem[\protect\citeauthoryear{{Lorimer}, {Bailes} \& {Harrison}}{{Lorimer}
  et~al.}{1997}]{lorimer1997}
{Lorimer} D.~R.,  {Bailes} M.,    {Harrison} P.~A.,  1997, \mnras, 289, 592

\bibitem[\protect\citeauthoryear{{Lorimer} et~al.,}{{Lorimer}
  et~al.}{2006}]{lorimer2006}
{Lorimer} D.~R.  et~al., 2006, \mnras, 372, 777

\bibitem[\protect\citeauthoryear{{Lyne} \& {Lorimer}}{{Lyne} \&
  {Lorimer}}{1994}]{lyne1994}
{Lyne} A.~G.,  {Lorimer} D.~R.,  1994, \nat, 369, 127

\bibitem[\protect\citeauthoryear{{Lyne} \& {Manchester}}{{Lyne} \&
  {Manchester}}{1988}]{lyne1988}
{Lyne} A.~G.,  {Manchester} R.~N.,  1988, \mnras, 234, 477

\bibitem[\protect\citeauthoryear{{Lyne}, {Ritchings} \& {Smith}}{{Lyne}
  et~al.}{1975}]{lyne1975}
{Lyne} A.~G.,  {Ritchings} R.~T.,    {Smith} F.~G.,  1975, \mnras, 171, 579

\bibitem[\protect\citeauthoryear{{Manchester}, {Fan}, {Lyne}, {Kaspi} \&
  {Crawford}}{{Manchester} et~al.}{2006}]{manchester2006}
{Manchester} R.~N.,  {Fan} G.,  {Lyne} A.~G.,  {Kaspi} V.~M.,    {Crawford} F.,
   2006, \apj, 649, 235

\bibitem[\protect\citeauthoryear{{McConnell}, {McCulloch}, {Hamilton}, {Ables},
  {Hall}, {Jacka} \& {Hunt}}{{McConnell} et~al.}{1991}]{mcconnell1991}
{McConnell} D.,  {McCulloch} P.~M.,  {Hamilton} P.~A.,  {Ables} J.~G.,  {Hall}
  P.~J.,  {Jacka} C.~E.,    {Hunt} A.~J.,  1991, \mnras, 249, 654

\bibitem[\protect\citeauthoryear{{Moe} \& {Di Stefano}}{{Moe} \& {Di
  Stefano}}{2017}]{moe2017}
{Moe} M.,  {Di Stefano} R.,  2017, \apjs, 230, 15

\bibitem[\protect\citeauthoryear{{Nelemans}, {Verbunt}, {Yungelson} \&
  {Portegies Zwart}}{{Nelemans} et~al.}{2000}]{nelemans2000}
{Nelemans} G.,  {Verbunt} F.,  {Yungelson} L.~R.,    {Portegies Zwart} S.~F.,
  2000, \aap, 360, 1011

\bibitem[\protect\citeauthoryear{{Nelemans}, {Yungelson}, {Portegies Zwart} \&
  {Verbunt}}{{Nelemans} et~al.}{2001}]{nelemans2001b}
{Nelemans} G.,  {Yungelson} L.~R.,  {Portegies Zwart} S.~F.,    {Verbunt} F.,
  2001, \aap, 365, 491

\bibitem[\protect\citeauthoryear{{Paczynski}}{{Paczynski}}{1976}]{paczynski1976}
{Paczynski} B.,  1976, in {Eggleton} P.,  {Mitton} S.,   {Whelan} J.,  eds,
  IAU Symposium Vol. 73, Structure and Evolution of Close Binary Systems. p.~75

\bibitem[\protect\citeauthoryear{{Pfahl}, {Rappaport}, {Podsiadlowski} \&
  {Spruit}}{{Pfahl} et~al.}{2002}]{pfahl2002}
{Pfahl} E.,  {Rappaport} S.,  {Podsiadlowski} P.,    {Spruit} H.,  2002, \apj,
  574, 364

\bibitem[\protect\citeauthoryear{{Podsiadlowski}, {Langer}, {Poelarends},
  {Rappaport}, {Heger} \& {Pfahl}}{{Podsiadlowski}
  et~al.}{2004}]{podsiadlowski2004}
{Podsiadlowski} P.,  {Langer} N.,  {Poelarends} A.~J.~T.,  {Rappaport} S.,
  {Heger} A.,    {Pfahl} E.,  2004, \apj, 612, 1044

\bibitem[\protect\citeauthoryear{{Portegies Zwart} \& {Verbunt}}{{Portegies
  Zwart} \& {Verbunt}}{1996}]{portegies1996}
{Portegies Zwart} S.~F.,  {Verbunt} F.,  1996, \aap, 309, 179

\bibitem[\protect\citeauthoryear{{Portegies Zwart} \& {Yungelson}}{{Portegies
  Zwart} \& {Yungelson}}{1998}]{portegies1998}
{Portegies Zwart} S.~F.,  {Yungelson} L.~R.,  1998, \aap, 332, 173

\bibitem[\protect\citeauthoryear{{Proszynski} \& {Przybycien}}{{Proszynski} \&
  {Przybycien}}{1984}]{proszynski1984}
{Proszynski} M.,  {Przybycien} D.,  1984, in {Reynolds} S.~P.,  {Stinebring}
  D.~R.,  eds, Birth and Evolution of Neutron Stars: Issues Raised by
  Millisecond Pulsars. p.~151

\bibitem[\protect\citeauthoryear{{Raghavan} et~al.,}{{Raghavan}
  et~al.}{2010}]{raghavan2010}
{Raghavan} D.  et~al., 2010, \apjs, 190, 1

\bibitem[\protect\citeauthoryear{{Ridley}, {Crawford}, {Lorimer}, {Bailey},
  {Madden}, {Anella} \& {Chennamangalam}}{{Ridley} et~al.}{2013}]{ridley2013}
{Ridley} J.~P.,  {Crawford} F.,  {Lorimer} D.~R.,  {Bailey} S.~R.,  {Madden}
  J.~H.,  {Anella} R.,    {Chennamangalam} J.,  2013, \mnras, 433, 138

\bibitem[\protect\citeauthoryear{{Ridley} \& {Lorimer}}{{Ridley} \&
  {Lorimer}}{2010}]{ridley2010}
{Ridley} J.~P.,  {Lorimer} D.~R.,  2010, \mnras, 406, L80

\bibitem[\protect\citeauthoryear{{Sana} et~al.,}{{Sana}
  et~al.}{2012}]{sana2012}
{Sana} H.  et~al., 2012, in {Drissen} L.,  {Robert} C.,  {St-Louis} N.,
  {Moffat} A.~F.~J.,  eds,  Astronomical Society of the Pacific Conference
  Series Vol. 465, Proceedings of a Scientific Meeting in Honor of Anthony F.
  J. Moffat. p.~284

\bibitem[\protect\citeauthoryear{{Sana} et~al.,}{{Sana}
  et~al.}{2014}]{sana2014}
{Sana} H.  et~al., 2014, \apjs, 215, 15

\bibitem[\protect\citeauthoryear{{Shtykovskiy} \& {Gilfanov}}{{Shtykovskiy} \&
  {Gilfanov}}{2005}]{shtykovskiy2005}
{Shtykovskiy} P.,  {Gilfanov} M.,  2005, \mnras, 362, 879

\bibitem[\protect\citeauthoryear{{Staveley-Smith} et~al.,}{{Staveley-Smith}
  et~al.}{1996}]{staveley1996}
{Staveley-Smith} L.  et~al., 1996, Publ. Astron. Soc. Australia, 13, 243

\bibitem[\protect\citeauthoryear{{Sturm} et~al.,}{{Sturm}
  et~al.}{2013}]{sturm2013}
{Sturm} R.  et~al., 2013, \aap, 558, A3

\bibitem[\protect\citeauthoryear{{Sturrock}}{{Sturrock}}{1971}]{sturrock1971}
{Sturrock} P.~A.,  1971, \apj, 164, 529

\bibitem[\protect\citeauthoryear{{Tauris}, {Langer} \&
  {Podsiadlowski}}{{Tauris} et~al.}{2015}]{tauris2015}
{Tauris} T.~M.,  {Langer} N.,    {Podsiadlowski} P.,  2015, \mnras, 451, 2123

\bibitem[\protect\citeauthoryear{{Tauris} \& {Manchester}}{{Tauris} \&
  {Manchester}}{1998}]{tauris1998}
{Tauris} T.~M.,  {Manchester} R.~N.,  1998, \mnras, 298, 625

\bibitem[\protect\citeauthoryear{{Titus} et~al.,}{{Titus}
  et~al.}{2019}]{titus2019}
{Titus} N.  et~al., 2019, \mnras, 487, 4332

\bibitem[\protect\citeauthoryear{{Toonen} \& {Nelemans}}{{Toonen} \&
  {Nelemans}}{2013}]{toonen2013}
{Toonen} S.,  {Nelemans} G.,  2013, \aap, 557, A87

\bibitem[\protect\citeauthoryear{{Toonen}, {Nelemans} \& {Portegies
  Zwart}}{{Toonen} et~al.}{2012}]{toonen2012}
{Toonen} S.,  {Nelemans} G.,    {Portegies Zwart} S.,  2012, \aap, 546, A70

\bibitem[\protect\citeauthoryear{{Tutukov} \& {Yungelson}}{{Tutukov} \&
  {Yungelson}}{1973}]{tutukov1973}
{Tutukov} A.,  {Yungelson} L.,  1973, Nauchnye Informatsii, 27, 70

\bibitem[\protect\citeauthoryear{{van den Heuvel}}{{van den
  Heuvel}}{2004}]{heuvel2004}
{van den Heuvel} E.~P.~J.,  2004, in {Schoenfelder} V.,  {Lichti} G.,
  {Winkler} C.,  eds,  ESA Special Publication Vol. 552, 5th INTEGRAL Workshop
  on the INTEGRAL Universe. p.~185

\bibitem[\protect\citeauthoryear{{van Haaften}, {Nelemans}, {Voss}, {van der
  Sluys} \& {Toonen}}{{van Haaften} et~al.}{2015}]{vanhaaften2015}
{van Haaften} L.~M.,  {Nelemans} G.,  {Voss} R.,  {van der Sluys} M.~V.,
  {Toonen} S.,  2015, \aap, 579, A33

\bibitem[\protect\citeauthoryear{{van Leeuwen} \& {Verbunt}}{{van Leeuwen} \&
  {Verbunt}}{2004}]{leeuwen2004}
{van Leeuwen} J.,  {Verbunt} F.,  2004, in {Camilo} F.,  {Gaensler} B.~M.,
  eds,  IAU Symposium Vol. 218, Young Neutron Stars and Their Environments.
  p.~41

\bibitem[\protect\citeauthoryear{{Verbunt}, {Igoshev} \& {Cator}}{{Verbunt}
  et~al.}{2017}]{verbunt2017}
{Verbunt} F.,  {Igoshev} A.,    {Cator} E.,  2017, \aap, 608, A57

\bibitem[\protect\citeauthoryear{{Vivekanand} \& {Narayan}}{{Vivekanand} \&
  {Narayan}}{1981}]{vivekanand1981}
{Vivekanand} M.,  {Narayan} R.,  1981, Journal of Astrophysics and Astronomy,
  2, 315

\bibitem[\protect\citeauthoryear{{Webbink}}{{Webbink}}{1984}]{webbink1984}
{Webbink} R.~F.,  1984, \apj, 277, 355

\bibitem[\protect\citeauthoryear{{Yokogawa}, {Imanishi}, {Tsujimoto},
  {Nishiuchi}, {Koyama}, {Nagase} \& {Corbet}}{{Yokogawa}
  et~al.}{2000}]{yokogawa2000}
{Yokogawa} J.,  {Imanishi} K.,  {Tsujimoto} M.,  {Nishiuchi} M.,  {Koyama} K.,
  {Nagase} F.,    {Corbet} R. H.~D.,  2000, \apjs, 128, 491

\bibitem[\protect\citeauthoryear{{Yoon}, {Woosley} \& {Langer}}{{Yoon}
  et~al.}{2010}]{yoon2010}
{Yoon} S.~C.,  {Woosley} S.~E.,    {Langer} N.,  2010, \apj, 725, 940

\bibitem[\protect\citeauthoryear{{Zaritsky}, {Harris}, {Thompson}, {Grebel} \&
  {Massey}}{{Zaritsky} et~al.}{2002}]{zaritsky2002}
{Zaritsky} D.,  {Harris} J.,  {Thompson} I.~B.,  {Grebel} E.~K.,    {Massey}
  P.,  2002, \aj, 123, 855

\bibitem[\protect\citeauthoryear{{Zorotovic}, {Schreiber}, {G{\"a}nsicke} \&
  {Nebot G{\'o}mez-Mor{\'a}n}}{{Zorotovic} et~al.}{2010}]{zorotovic2010}
{Zorotovic} M.,  {Schreiber} M.~R.,  {G{\"a}nsicke} B.~T.,    {Nebot
  G{\'o}mez-Mor{\'a}n} A.,  2010, \aap, 520, A86

\bibitem[\protect\citeauthoryear{{Zorotovic}, {Schreiber}, {Garc{\'\i}a-Berro},
  {Camacho}, {Torres}, {Rebassa-Mansergas} \& {G{\"a}nsicke}}{{Zorotovic}
  et~al.}{2014}]{zorotovic2014}
{Zorotovic} M.,  {Schreiber} M.~R.,  {Garc{\'\i}a-Berro} E.,  {Camacho} J.,
  {Torres} S.,  {Rebassa-Mansergas} A.,    {G{\"a}nsicke} B.~T.,  2014, \aap,
  568, A68

\end{thebibliography}


\appendix

\section{Simulation tables}\label{sec:tables}

Here follows the results of all the simulations where the \citet{antoniou2010} SFH was assumed.

\begin{table*}
\centering
\caption{Simulation results for the \citet{tauris1998} beaming fraction, and $\sigma_{\text{Lcorr}}$\,=\,0.8 radio luminosity model.}
\label{tab:TM199808}
\begin{tabular}{lllllllllllll}
\hline
Model & \multicolumn{2}{|c|}{Total Pulsars} & \multicolumn{2}{|c|}{Detached} & \multicolumn{2}{|c|}{Disrupted} & \multicolumn{2}{|c|}{\citet{crawford2001}} & \multicolumn{2}{|c|}{\citet{manchester2006}} & \multicolumn{2}{|c|}{MeerKAT}\\
CE & $\gamma\alpha$ & $\alpha\alpha$ & $\gamma\alpha$& $\alpha\alpha$ & $\gamma\alpha$& $\alpha\alpha$ & $\gamma\alpha$& $\alpha\alpha$ & $\gamma\alpha$& $\alpha\alpha$ & $\gamma\alpha$& $\alpha\alpha$\\
\hline
\hline
V4CE2  &  14889  &  14866  &  5  &  5  &  95  &  95  &  4.08  &  4.00  &  4.05  &  3.96  &  17.30  &  17.26 \\
H4CE2  &  15234  &  15082  &  2  &  2  &  98  &  98  &  3.54  &  3.58  &  3.49  &  3.57  &  16.57  &  16.36 \\
V4CE25  &  14717  &  14053  &  4  &  4  &  96  &  96  &  4.83  &  4.57  &  4.80  &  4.53  &  19.61  &  18.81 \\
H4CE25  &  15005  &  14227  &  2  &  2  &  98  &  98  &  4.72  &  4.32  &  4.69  &  4.28  &  18.93  &  18.11 \\
0.1V4CE2  &  13707  &  13395  &  22  &  22  &  78  &  78  &  5.33  &  5.21  &  5.28  &  5.19  &  21.18  &  19.66 \\
0.1V4CE25  &  13374  &  12439  &  19  &  17  &  81  &  83  &  5.78  &  5.20  &  5.73  &  5.16  &  22.84  &  20.73 \\
V8CE2  &  15071  &  14693  &  5  &  5  &  95  &  95  &  4.32  &  3.97  &  4.28  &  3.95  &  18.48  &  17.58 \\
H8CE2  &  15247  &  14893  &  2  &  2  &  98  &  98  &  3.81  &  3.54  &  3.78  &  3.50  &  17.35  &  16.92 \\
V8CE25  &  14468  &  13747  &  4  &  4  &  96  &  96  &  4.09  &  3.66  &  4.06  &  3.63  &  16.22  &  15.00 \\
H8CE25  &  14691  &  13933  &  2  &  2  &  98  &  98  &  3.74  &  3.38  &  3.72  &  3.35  &  15.76  &  13.93 \\
0.1V8CE2  &  12957  &  13303  &  20  &  23  &  80  &  77  &  5.29  &  5.69  &  5.26  &  5.66  &  19.74  &  22.06 \\
0.1V8CE25  &  13484  &  12092  &  23  &  17  &  77  &  83  &  6.01  &  4.48  &  5.97  &  4.44  &  23.04  &  17.11 \\
\hline
\end{tabular}
\end{table*}

\begin{table*}
\centering
\caption{Simulation results for the \citet{tauris1998} beaming fraction, and $\sigma_{\text{Lcorr}}$\,=\,2.0 radio luminosity model.}
\label{tab:TM199820}
\begin{tabular}{lllllllllllll}
\hline
Model & \multicolumn{2}{|c|}{Total Pulsars} & \multicolumn{2}{|c|}{Detached} & \multicolumn{2}{|c|}{Disrupted} & \multicolumn{2}{|c|}{\citet{crawford2001}} & \multicolumn{2}{|c|}{\citet{manchester2006}} & \multicolumn{2}{|c|}{MeerKAT}\\
CE & $\gamma\alpha$ & $\alpha\alpha$ & $\gamma\alpha$& $\alpha\alpha$ & $\gamma\alpha$& $\alpha\alpha$ & $\gamma\alpha$& $\alpha\alpha$ & $\gamma\alpha$& $\alpha\alpha$ & $\gamma\alpha$& $\alpha\alpha$\\
\hline
\hline
V4CE2  &  14889  &  14866  &  5  &  5  &  95  &  95  &  14.25  &  14.23  &  14.16  &  14.12  &  111.11  &  111.36 \\
H4CE2  &  15234  &  15082  &  2  &  2  &  98  &  98  &  13.52  &  13.42  &  13.40  &  13.27  &  111.49  &  110.33 \\
V4CE25  &  14717  &  14053  &  4  &  4  &  96  &  96  &  16.10  &  15.34  &  15.95  &  15.23  &  112.64  &  111.23 \\
H4CE25  &  15005  &  14227  &  2  &  2  &  98  &  98  &  15.61  &  14.64  &  15.51  &  14.54  &  110.86  &  109.06 \\
0.1V4CE2  &  13707  &  13395  &  22  &  22  &  78  &  78  &  17.68  &  16.62  &  17.56  &  16.51  &  112.86  &  107.74 \\
0.1V4CE25  &  13374  &  12439  &  19  &  17  &  81  &  83  &  19.00  &  16.93  &  18.88  &  16.81  &  114.79  &  111.07 \\
V8CE2  &  15071  &  14693  &  5  &  5  &  95  &  95  &  15.28  &  14.40  &  15.17  &  14.30  &  109.72  &  107.25 \\
H8CE2  &  15247  &  14893  &  2  &  2  &  98  &  98  &  14.32  &  13.86  &  14.22  &  13.73  &  106.81  &  104.70 \\
V8CE25  &  14468  &  13747  &  4  &  4  &  96  &  96  &  13.47  &  12.18  &  13.34  &  12.09  &  100.64  &  97.30 \\
H8CE25  &  14691  &  13933  &  2  &  2  &  98  &  98  &  12.87  &  11.38  &  12.77  &  11.29  &  100.17  &  94.39 \\
0.1V8CE2  &  12957  &  13303  &  20  &  23  &  80  &  77  &  16.65  &  18.64  &  16.54  &  18.54  &  104.04  &  106.21 \\
0.1V8CE25  &  13484  &  12092  &  23  &  17  &  77  &  83  &  19.45  &  14.21  &  19.35  &  14.13  &  109.88  &  97.01 \\

\hline
\end{tabular}
\end{table*}

\begin{table*}
\centering
\caption{Simulation results for the \citet{lyne1988} beaming fraction, and $\sigma_{\text{Lcorr}}$\,=\,0.8 radio luminosity model.}
\label{tab:LM198808}
\begin{tabular}{lllllllllllll}
\hline
Model & \multicolumn{2}{|c|}{Total Pulsars} & \multicolumn{2}{|c|}{Detached} & \multicolumn{2}{|c|}{Disrupted} & \multicolumn{2}{|c|}{\citet{crawford2001}} & \multicolumn{2}{|c|}{\citet{manchester2006}} & \multicolumn{2}{|c|}{MeerKAT}\\
CE & $\gamma\alpha$ & $\alpha\alpha$ & $\gamma\alpha$& $\alpha\alpha$ & $\gamma\alpha$& $\alpha\alpha$ & $\gamma\alpha$& $\alpha\alpha$ & $\gamma\alpha$& $\alpha\alpha$ & $\gamma\alpha$& $\alpha\alpha$\\
\hline
\hline
V4CE2  &  21365  &  21323  &  5  &  5  &  95  &  95  &  3.44  &  3.38  &  3.41  &  3.35  &  15.33  &  15.29 \\
H4CE2  &  21861  &  21635  &  2  &  2  &  98  &  98  &  2.99  &  3.02  &  2.95  &  3.01  &  14.70  &  14.51 \\
V4CE25  &  21116  &  20159  &  4  &  4  &  96  &  96  &  4.08  &  3.86  &  4.05  &  3.83  &  17.29  &  16.55 \\
H4CE25  &  21538  &  20422  &  2  &  2  &  98  &  98  &  3.98  &  3.64  &  3.96  &  3.62  &  16.67  &  15.92 \\
0.1V4CE2  &  19671  &  19219  &  22  &  22  &  78  &  78  &  4.52  &  4.42  &  4.48  &  4.40  &  18.82  &  17.45 \\
0.1V4CE25  &  19155  &  17802  &  19  &  16  &  81  &  84  &  4.88  &  4.39  &  4.84  &  4.36  &  20.15  &  18.23 \\
V8CE2  &  21676  &  21122  &  5  &  5  &  95  &  95  &  3.64  &  3.34  &  3.61  &  3.33  &  16.40  &  15.62 \\
H8CE2  &  21939  &  21422  &  2  &  2  &  98  &  98  &  3.21  &  3.00  &  3.18  &  2.96  &  15.39  &  15.04 \\
V8CE25  &  20800  &  19777  &  4  &  4  &  96  &  96  &  3.46  &  3.10  &  3.44  &  3.08  &  14.34  &  13.24 \\
H8CE25  &  21137  &  20052  &  2  &  2  &  98  &  98  &  3.16  &  2.86  &  3.14  &  2.84  &  13.92  &  12.30 \\
0.1V8CE2  &  18596  &  19121  &  19  &  23  &  81  &  77  &  4.48  &  4.82  &  4.45  &  4.79  &  17.46  &  19.60 \\
0.1V8CE25  &  19395  &  17362  &  23  &  17  &  77  &  83  &  5.10  &  3.81  &  5.07  &  3.77  &  20.51  &  15.11 \\
\hline
\end{tabular}
\end{table*}

\begin{table*}
\centering
\caption{Simulation results for the \citet{lyne1988} beaming fraction, and $\sigma_{\text{Lcorr}}$\,=\,2.0 radio luminosity model.}
\label{tab:LM198820}
\begin{tabular}{lllllllllllll}
\hline
Model & \multicolumn{2}{|c|}{Total Pulsars} & \multicolumn{2}{|c|}{Detached} & \multicolumn{2}{|c|}{Disrupted} & \multicolumn{2}{|c|}{\citet{crawford2001}} & \multicolumn{2}{|c|}{\citet{manchester2006}} & \multicolumn{2}{|c|}{MeerKAT}\\
CE & $\gamma\alpha$ & $\alpha\alpha$ & $\gamma\alpha$& $\alpha\alpha$ & $\gamma\alpha$& $\alpha\alpha$ & $\gamma\alpha$& $\alpha\alpha$ & $\gamma\alpha$& $\alpha\alpha$ & $\gamma\alpha$& $\alpha\alpha$\\
\hline
\hline
V4CE2  &  21365  &  21323  &  5  &  5  &  95  &  95  &  12.63  &  12.61  &  12.55  &  12.52  &  97.34  &  97.51 \\
H4CE2  &  21861  &  21635  &  2  &  2  &  98  &  98  &  12.00  &  11.90  &  11.90  &  11.77  &  97.63  &  96.63 \\
V4CE25  &  21116  &  20159  &  4  &  4  &  96  &  96  &  14.15  &  13.46  &  14.03  &  13.37  &  98.66  &  97.81 \\
H4CE25  &  21538  &  20422  &  2  &  2  &  98  &  98  &  13.71  &  12.86  &  13.62  &  12.77  &  97.03  &  95.74 \\
0.1V4CE2  &  19671  &  19219  &  22  &  22  &  78  &  78  &  15.72  &  14.74  &  15.63  &  14.65  &  99.39  &  94.86 \\
0.1V4CE25  &  19155  &  17802  &  19  &  16  &  81  &  84  &  16.72  &  14.86  &  16.61  &  14.76  &  100.99  &  97.98 \\
V8CE2  &  21676  &  21122  &  5  &  5  &  95  &  95  &  13.55  &  12.78  &  13.46  &  12.70  &  96.36  &  94.19 \\
H8CE2  &  21939  &  21422  &  2  &  2  &  98  &  98  &  12.69  &  12.30  &  12.61  &  12.20  &  93.82  &  91.93 \\
V8CE25  &  20800  &  19777  &  4  &  4  &  96  &  96  &  11.91  &  10.75  &  11.80  &  10.67  &  88.25  &  85.43 \\
H8CE25  &  21137  &  20052  &  2  &  2  &  98  &  98  &  11.36  &  10.04  &  11.28  &  9.97  &  87.71  &  82.85 \\
0.1V8CE2  &  18596  &  19121  &  19  &  23  &  81  &  77  &  14.71  &  16.56  &  14.62  &  16.47  &  91.59  &  93.80 \\
0.1V8CE25  &  19395  &  17362  &  23  &  17  &  77  &  83  &  17.29  &  12.56  &  17.21  &  12.49  &  97.13  &  85.52 \\
\hline
\end{tabular}
\end{table*}

\bsp	
\label{lastpage}
\end{document}